\documentclass[acmsmall]{acmart}

\AtBeginDocument{%
  \providecommand\BibTeX{{%
    \normalfont B\kern-0.5em{\scshape i\kern-0.25em b}\kern-0.8em\TeX}}}

\setcopyright{acmlicensed}
\acmJournal{PACMHCI}
\acmYear{2023} \acmVolume{7} \acmNumber{CSCW1}
\acmArticle{129} \acmMonth{4} \acmPrice{15.00}
\acmDOI{10.1145/3579605}
\acmPrice{15.00}
\received{July 2022}
\received[revised]{October 2022}
\received[accepted]{January 2023}

\usepackage{enumitem}
\usepackage{xspace}
\usepackage{subfigure}
\usepackage[normalem]{ulem}
\usepackage{footnote}

\makesavenoteenv{tabular}
\definecolor{AIColor}{rgb}{0.0, 0.5, 0.0}
\definecolor{TaskColor}{rgb}{0.85, 0.65, 0.13}
\definecolor{BonusColor}{rgb}{0.6, 0.4, 0.8}

\makeatletter
\DeclareRobustCommand\onedot{\futurelet\@let@token\@onedot}
\def\@onedot{\ifx\@let@token.\else.\null\fi\xspace}

\def\eg{e.g\onedot} 
\def\ie{i.e\onedot}

\makeatother

\newcommand{\edit}[1]{{
{#1}
}}

\usepackage[bottom]{footmisc} % makes sure that footnotes are at the bottom 

\begin{document}

\title[Explanations Reduce Overreliance]{Explanations \underline{Can} Reduce Overreliance on AI Systems During Decision-Making}
%When Do XAI Methods Work? A Cost-Benefit Approach to Human-AI Collaboration

\author{Helena Vasconcelos}
\email{helenav@cs.stanford.edu}
\affiliation{%
  \institution{Stanford University}
  % \city{Stanford}
  % \state{California}
  \country{USA}
}

\author{Matthew J{\"o}rke}
\email{joerke@cs.stanford.edu}
\affiliation{%
  \institution{Stanford University}
  % \city{Stanford}
  % \state{California}
  \country{USA}
}

\author{Madeleine Grunde-McLaughlin}
\email{mgrunde@cs.washington.edu}
\affiliation{%
  \institution{University of Washington}
  % \city{Seattle}
  % \state{Washington}
  \country{USA}
}

\author{Tobias Gerstenberg}
\email{gerstenberg@stanford.edu}
\affiliation{%
  \institution{Stanford University}
  % \city{Stanford}
  % \state{California}
  \country{USA}
}

\author{Michael S. Bernstein}
\email{msb@cs.stanford.edu}
\affiliation{%
  \institution{Stanford University}
  % \city{Stanford}
  % \state{California}
  \country{USA}
}

\author{Ranjay Krishna}
\email{ranjay@cs.washington.edu}
\affiliation{%
  \institution{University of Washington}
  % \city{Seattle}
  % \state{Washington}
  \country{USA}
}

% \author{Anonymized for submission}

\renewcommand{\shortauthors}{Vasconcelos, et al.}
% \renewcommand{\shortauthors}{Anonymous}
\begin{abstract}
Prior work has identified a resilient phenomenon that threatens the performance of human-AI decision-making teams: \textit{overreliance}, when people agree with an AI, even when it is incorrect. Surprisingly, overreliance does not reduce when the AI produces explanations for its predictions, compared to only providing predictions. Some have argued that overreliance results from cognitive biases or uncalibrated trust, attributing overreliance to an inevitability of human cognition. By contrast, our paper argues that people strategically choose whether or not to engage with an AI explanation, demonstrating empirically that there \textit{are} scenarios where AI explanations reduce overreliance. To achieve this, we formalize this strategic choice in a cost-benefit framework, where the costs and benefits of engaging with the task are weighed against the costs and benefits of relying on the AI. We manipulate the costs and benefits in a maze task, where participants collaborate with a simulated AI to find the exit of a maze.
Through $5$ studies $(N = 731)$, we find that costs such as task difficulty (Study 1), explanation difficulty (Study 2, 3), and benefits such as monetary compensation (Study 4) affect overreliance.
Finally, Study 5 adapts the Cognitive Effort Discounting paradigm to quantify the utility of different explanations, providing further support for our framework.
Our results suggest that some of the null effects found in literature could be due in part to the explanation not sufficiently reducing the costs of verifying the AI's prediction.

\end{abstract}





\maketitle

\section{Introduction}

\begin{figure}[tb]
    \centering
    \includegraphics[width=0.65\textwidth]{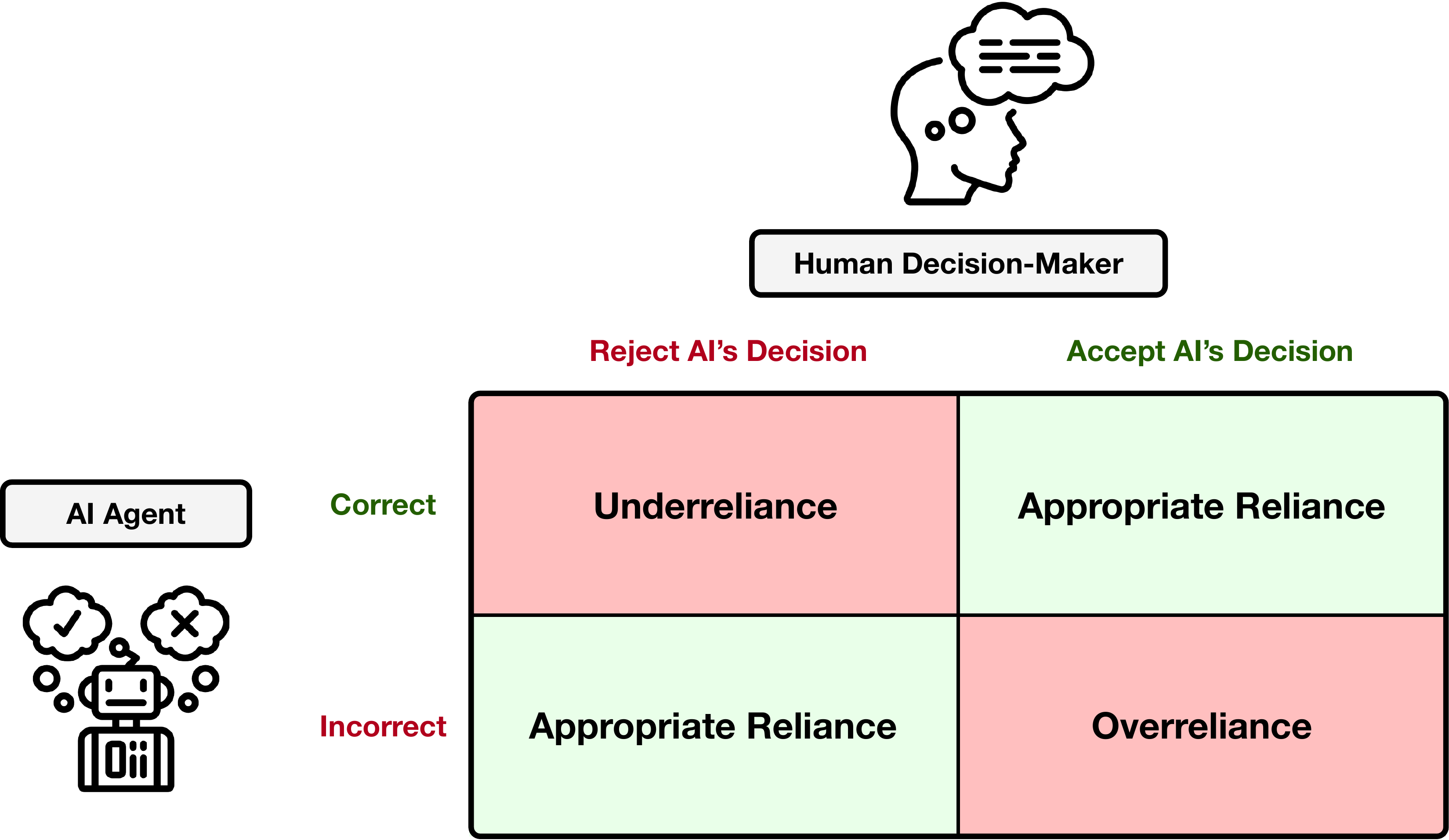}
    \caption{A simplified diagram outlining the different ways people rely on AI systems. Provided with a prediction from an AI system, a human decision-maker has the choice to either accept or reject the AI's prediction. Appropriate reliance occurs when the human accepts a correct AI prediction or corrects an incorrect AI prediction. Underreliance occurs when the human fails to accept an correct AI prediction. Overreliance occurs when the human fails to correct an incorrect AI prediction.}
    \label{fig:reliance-matrix}
\end{figure}

Human-AI decision-making has fallen short of its goal of achieving complementarity: Human-AI teams ought to perform better, and therefore make better decisions, than either humans or AIs alone~\cite{bansal2021does}. Unfortunately, the literature has repeatedly established that people are influenced by the AI and often accept incorrect decisions without verifying whether the AI was correct. This phenomenon is known as \textit{overreliance}~\cite{buccinca2021trust}. Of the possible error types in human-AI decision-making (Figure~\ref{fig:reliance-matrix}), overreliance is the most frequent outcome in empirical AI studies~\cite{bansal2021does,bucina2020proxy,lai2020chicago,zhang2020effect,buccinca2021trust}. Overreliance can be seen as foregoing agency and accountability to the AI when making final decisions~\cite{lai2019human,jacobs2021machine,bussone2015role}, and is of particular concern in high-stakes domains, running the risk of reinforcing machine bias~\cite{green2019principles,angwin2016machine} under the guise of human agency. 

Explainable AI (XAI) has been proposed as a mitigation, but experimental evidence has so far failed to find support that explanations reduce overreliance. The intuition runs that, if people see an explanation that is incorrect, they will more carefully scrutinize the prediction and be less likely to overrely~\cite{bayati2014data,bussone2015role,caruana2015intelligible}. However, empirically, explanations have failed to improve levels of overreliance~\cite{bansal2021does,buccinca2021trust}. Potential cognitive mechanisms that underlie these results suggest that the mere presence of an explanation increases trust~\cite{yu2019trust,zhang2020effect}, and that explanations anchor humans to the prediction~\cite{wang2019theory,buccinca2021trust}. Scaffolds such as cognitive forcing functions, which force people to engage more thoughtfully with the AI's explanations, can reduce overeliance~\cite{buccinca2021trust}. Taken together, existing literature puts forth a resilient finding: that explanations exacerbate or do not alter overreliance unless intervention methods in the form of forcing functions compel people to engage with the task.

In this paper, we produce evidence that explanations can in fact reduce overreliance without the need for forcing functions.
To demonstrate this, we begin by observing that real-world tasks and explanations can vary greatly in their cognitive costs~\cite{buccinca2021trust}: tasks can be cognitively costly (difficult) to complete, or cognitively cheap (easy); checking explanations can likewise be cognitively costly or cheap.
We argue that the null effects found in the literature---that explanations do not reduce overreliance---occur because prior work has focused on a narrow slice of this cost space where these two costs are nearly equal and both nontrivially costly.
In this situation, where both the task and the explanation are costly, we might expect people to just rely on the AI's prediction without carefully checking it. 
For example, answering a reading comprehension question requires reading a passage and incurs a similar cognitive effort as reading the AI generated natural language explanation~\cite{bansal2021does}; similarly, guessing the calories in a photo of food attracts a similar cost as guessing and summing up the calories of the individual ingredients produced as an AI explanation~\cite{buccinca2021trust}.
If it takes roughly as much effort to verify an AI's explanation as it does to complete the task manually, then we should expect people to sometimes forego verifying the explanation (either by overrelying or by completing the task alone).
But, if we were to make the task much more challenging and thus more costly, while the explanation remains equally costly as it was before, the explanation would suddenly become strongly preferable and overreliance would reduce. This is because, relatively speaking, verifying the AI's explanation has become very easy to do, compared to doing the task alone (very costly) or relying on the AI (undesirable because of possible errors). Likewise, if the explanation were to become substantially easier to check than it was before, it would become preferable and overreliance would reduce.

We introduce a theoretical framework to reason about the conditions under which explanations will reduce overreliance. We then use this framework to predict new conditions under which explanations will in fact reduce overreliance. 
Our framework, in detail, is a cost-benefit analysis wherein people strategically weight the cognitive effort required to complete the task alone or verify the AI against the effortless strategy of overreliance.
It suggests that when tasks require more effort and the explanations sufficiently reduce the cost of verifying with the AI, or when the explanations are easier to understand, explanations will reduce overreliance. 

We test this framework through a series of five experiments $(N=731)$ with a simulated AI in a maze-solving task. Study 1 $(N = 340)$ manipulates task difficulty. It replicates prior results that receiving an AI's prediction versus receiving an AI's prediction with an explanation have equal levels of overreliance—but only for an easy task. As the task becomes more effortful, and engaging with the explanation becomes less relative cognitive effort, XAI overreliance becomes lesser than AI overreliance.
Study 2 $(N = 340)$ and Study 3 $(N = 286)$ manipulate the cognitive cost of understanding and engaging with the AI's explanations, and find, again, that when the cognitive effort is lower, the reduction in overreliance is larger.
Study 4 $(N = 114)$ tests whether the cognitive effort (costs) are balanced against the incentives (benefits) accrued for expending those costs---the benefit piece of our cost-benefit framework---by manipulating the financial incentive for a correct answer. We verify that, as the monetary reward for accuracy increases, overreliance decreases.
Study 5 $(N = 76)$ adapts the Cognitive Effort Discounting paradigm~\cite{westbrook2013subjective} to synthesize the costs (Studies 1-3) with benefits (Study 4) through the lens of economic utility. This final study measures the financial value (utility) attached to explanations and validates our hypotheses, finding that people attach more utility to explanations that are easier to understand and attach more utility to explanations in harder tasks.

Our results indicate that overreliance is not an inevitability of cognition but a strategic decision where people are responsive to the costs and benefits. Beyond the immediate results of our study, our work suggests that designers of collaborative human-AI systems should carefully consider how explanations reduce the cognitive effort of verifying an AI in comparison to a low-effort strategy of overreliance. 
\section{Background}

Empirical HCI studies of human-AI decision-making find little evidence that human-AI teams actually achieve complementary performance in which teams perform better than a human or an AI alone, even when explanations are provided~\cite{bansal2021does}.
Since AIs outperform humans in many task domains, a strategy which blindly relies on model predictions is likely to achieve similar performance to one which utilizes explanations to evaluate model outputs. 
Human-AI performance generally only exceeds human-only performance~\cite{feng2019ai,lai2019human,bucina2020proxy,lundberg2018explainable,green2019principles,poursabzisangdeh2021manipulating,zhang2020effect}; 
stark performance differences between humans and AIs make it such that one would achieve the highest performance by removing humans from the loop entirely.
Therefore, among studies that report that performance improves when an AI gives explanations~\cite{lai2019human,lai2020chicago,bucina2020proxy,horne2019rating}, it is unclear whether explanations truly improve understanding of model capabilities, or if they instead serve as a signal for blind trust in a AI that is more accurate than the human baseline~\cite{bansal2021does,buccinca2021trust,eiband2019impact}. 

In light of this conundrum, recent studies evaluate human-AI decision-making in domains with roughly equal human and AI performance. Unfortunately, they do not find that explanations improve performance above a prediction-only baseline~\cite{chu2020visual,bansal2021does}, offering evidence for the latter theory—that explanations exacerbate overreliance. In such complementary domains, explanations (even those generated by humans~\cite{bansal2021does}) appear to increase reliance on both correct and incorrect AI predictions. Among inputs incorrectly predicted by the AI, humans achieve worse performance than if they had completed the task without any AI partner. In light of growing concerns surrounding machine bias~\cite{green2019principles,angwin2016machine}, such \textit{overreliance} on incorrect model predictions is particularly concerning and is the primary focus of our work. We investigate the conditions under which people engage in effortful thinking to verify AI predictions, reducing overreliance.
We are guided by the following research question:

\begin{quote}
\textsc{Research Question}: \textit{Under what conditions do people engage with explanations to verify AI predictions, reducing overreliance?}
\end{quote}

%-------------------------------------------------------------------------------
\subsection{Cognitive biases in decision-making}
Research to this point focuses on elucidating the cognitive mechanisms and biases that cause overreliance. 
For decades, cognitive psychology literature on mental effort and advice utilization has attempted to understand the cognitive processes involved in decision-making. 
Pioneered by Tversky \& Kahneman~\cite{kahneman2003perspective,kahneman2011thinking}, \textit{dual process theory} accounts for reasoning by means of two distinct cognitive subsystems: one that is unconscious, automatic, fast, and intuitive (System~1) and one that is deliberate, effortful, slow, and analytical (System~2). System 1 processes employ heuristics to arrive at fast and efficient conclusions, without which all thinking would be slow and arduous. While System~1 thinking can greatly speed up decision-making, its reliance on cognitive shortcuts is also the source of many known cognitive biases~\cite{kahneman2011thinking}. 

Systems 1 and 2 can interact with and override each other throughout decision-making processes, leading to judgements that are often hybrids of intuitive and analytical thinking~\cite{croskerry2009universal}. 
As a result, cognitive biases are pervasive in most forms of decision-making, even in high-stakes tasks with domain experts. For example, the role of cognitive biases is well-documented in medicine~\cite{lambe2016dual,croskerry2009clinical,croskerry2009universal}, with survey studies indicating a diagnostic error rate of around 10-15\%, with 75\% of these errors resulting from cognitive factors. Other work has investigated the role of cognitive biases in foreign policy~\cite{lebovic2013policy}, the initiation of war~\cite{vaughn2016war}, judicial sentencing decisions~\cite{guthrie2007blinking}, national intelligence~\cite{mandel2018correcting}, or catastrophe risk assessment~\cite{vasiljevic2013reasoning}. 
% \matt{These previous citations were all taken from~\cite{croskerry2017model}, maybe we should just cite that.} 
These cognitive biases manifest themselves in the way people process information during decision-making.

Based on dual process theory, a recent study investigates the use of cognitive forcing functions—interventions designed to encourage effortful, analytical thinking \edit{(e.g., forcing the user to wait for a certain period of time before they receive an explanation from the AI)}—as a means for reducing overreliance~\cite{buccinca2021trust}. The authors find that forcing functions successfully reduce overreliance on incorrect model predictions. However, they also reduce reliance on correct model predictions, yielding no significant differences when humans are not aided by an AI. 
The authors also identified an interesting trade-off: participants performed best in the conditions they preferred and trusted the least. 
We provide another lens through which to view this line of work: one that uses a cost-benefit framework where cognitive forcing functions can be viewed as manipulations for cost, i.e.~by necessitating cognitive effort and analytical thinking. 

%-------------------------------------------------------------------------------
\subsection{Decision-making in behavioral economics}
Dual process theory makes clear that System 2 can override default biases and heuristics~\cite{evans2003two}.
However, human decision-makers are ``satisficers'', devoting only as much effort as is minimally satisfactory rather than aiming for optimal decision-making outcomes~\cite{simon1956rational}. 
To frame this aversion to engaging in effortful thinking in economic terms:
people weigh the potential benefits of cognitive effort against its perceived costs~\cite{navon1977economy,kool2018mental}. This economic framing of cognitive effort allocation is supported by evidence from behavioral experiments~\cite{kool2010decision,westbrook2013subjective,kool2014labor}, neuroscientific studies~\cite{mcguire2010prefrontal,botvinick2014computational,shenhav2017toward}, and computer simulations~\cite{johnson1985effort,payne1988adaptive}. In this economic framing, effort-based decision-making is modeled as a selection among different cognitive strategies, each associated with its own perceived benefits (in terms of task performance) and costs (in terms of cognitive effort). Decision-makers navigate this effort-accuracy trade-off~\cite{johnson1985effort} by adopting the strategy which maximizes subjective utility as a function of costs and benefits~\cite{kool2018mental}. In the context of computer-based decision aids, prior work has found that decision aids which reduced the cognitive effort associated with a particular strategy induced behavior associated with that strategy~\cite{todd1994influence}, implying that decision-aid designers can influence cognitive strategy selection by manipulating costs. Similar studies~\cite{kool2017cost,sandra2018cognitive} also provide support for the influence of benefits: when higher rewards are at stake, people engage in more effortful strategies.
Our study investigates the effect of XAI methods by manipulating the costs associated with a high-effort strategy which verifies an AI's predictions versus the low-effort strategy of overreliance.

\section{Framework}

\begin{figure}[t]
    \centering
    \includegraphics[width=0.65\textwidth]{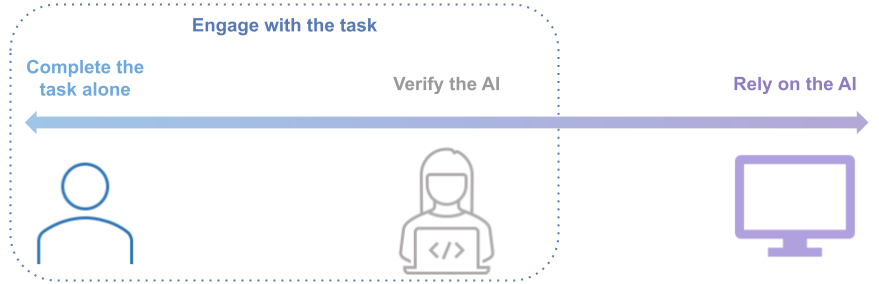}
    \caption{We define two prototype strategies: (1) engage with the task (in the form of completing the task alone or verifying the AI's prediction and/or explanation), and (2) rely on the AI. Of course, decision-makers can lie on this continuum, especially as they interact with AI over long periods of time, but these strategies help ground our framework.}
    \label{fig:Strategy-Figure}
\end{figure}

Theories proposed in prior work—from anchoring biases~\cite{wang2019theory,buccinca2021trust} to humans' innate adversity to cognitive effort~\cite{navon1977economy,kool2018mental,zhang2020effect,buccinca2021trust}—suggest that overreliance, or the frequency with which people agree with an AI when it is incorrect~\cite{buccinca2021trust}, is the default and inevitable state of human performance when presented with an explanation.
However, clearly sometimes people do engage with automated systems AIs in calibrated ways. For example, people rarely choose a suggested auto-reply to an important email to their boss~\cite{hancock2020ai}, and will override AI suggestions if a virtual assistant misunderstands a request~\cite{luger2016like}. Furthermore, these frameworks do not account for research in cognitive psychology in which explanations improve performance in human-human contexts~\cite{zuboff1988age,barber1983logic,kulesza2013too,lombrozo2006structure,chen2019cicero}.

To account for this contradiction, we propose that overreliance is not an inevitability of cognition, but instead the result of a strategic choice. For example, a person may be less likely to overrely when the \textit{benefit} of getting the right answer is high (e.g., not sending an unprofessional email to a boss). Alternatively, a person might be less likely to overrely when the \textit{cost} associated with extracting the right answer is low (e.g., deciding whether a tweet has positive sentiment).

Prior work in behavioral economics uses cost-benefit frameworks~\cite{navon1977economy,kool2018mental} to explain which decision making strategies people choose. In this framework, people subconsciously weigh the costs (e.g., cognitive effort, time) and benefits (e.g., task performance, rewards) of performing different cognitive strategies. The person forms an opinion on the value of each strategy, called the \textit{subjective utility} of that strategy, by judging its ability to minimize costs and maximize benefits. They enact the strategy that they value to have the highest subjective utility. Utility if often defined to a function of perceived costs and benefits.

We propose that this cost-benefit framework from behavioral economics also applies to the way people collaborate with AI predictions and explanations. In the context of human-AI decision making, this framework suggests that people compare the potential benefits of engaging with the task (\ie professional accomplishment, monetary reward) weighed against its inherent costs (\ie cognitive effort, time)\footnote{\edit{We note that there are effects of other factors, like trust and intrinsic rewards, but we do not explore these in this paper. See: Section 10.4.}}. People are likely to overrely if engaging with the task is not the optimal strategy. When people choose to engage with the task, they are adopting one of two strategies: either they do the task alone, ignoring any AI support, or they verify the AI's prediction, sometimes using AI generated explanations if they are present.

Perceptions of costs in the form of cognitive effort, time, etc., as well as benefits in the form of monetary reward, stakes, etc. are subject to individual differences~\cite{kool2010decision,tversky1992advances}. However, we predict that the following principle holds: \textit{if an element of a task (e.g., an explanation) reduces the cost (via reducing the difficulty of the task or other means) and/or increases the benefits (via increasing monetary reward or other means) of a particular decision-making strategy relative to other strategies, a decision-maker will be more inclined to employ that strategy.}\footnote{This principle is adapted from~\cite[p.~50-51]{todd1994influence}.} Therefore, when aggregating across many decision-makers, we expect that conditions which support a particular strategy will induce behavior consistent with that strategy.

This framework offers a possible explanation for why XAIs have not decreased overreliance compared to traditional AIs in prior work: we have been focused on a narrow slice of the cost-benefit space. In particular, receiving the AI's explanation does not reduce the costs of engaging with the task. This is to say that, in these cases, the cost of engaging with the task using the AI's explanation is roughly equal to engaging via using the prediction or by completing the task alone. In this case, we do not expect to see differences between overreliance levels with explanations, since there is no difference in cost. Therefore, the strategy to engage with the task does not have a change in utility.

However, other parts of the cost-benefit space would predict different outcomes. Consider the following example: if the task were more difficult and if the AI's explanation reduced the cost of engaging with the task, the strategy to engage with the task would seem relatively much more attractive. This is because relying remains undesirable due to the prospect of the AI making mistakes (lowering the benefits of this strategy) and engaging via completing the task alone remains very costly. In this case, engaging with the task (likely by engaging with the explanations) is desirable and we can moderately assume that overreliance will decrease. 

\begin{figure}[t]
    \centering
    \includegraphics[width=0.75\textwidth]{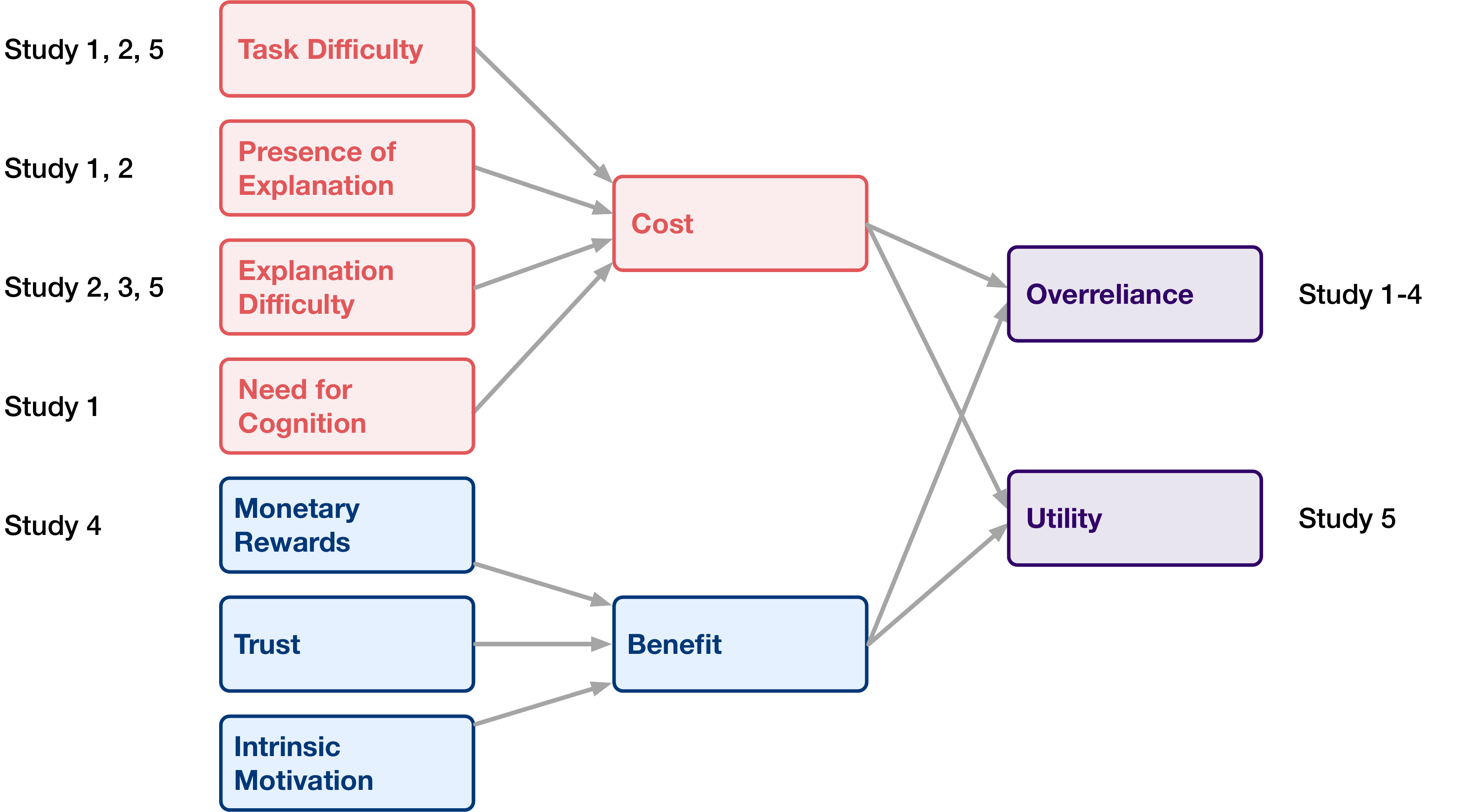}
    \caption{Outline of some of the factors that affect the costs and benefits of engaging with the explanations, as well as our associated studies that manipulate or measure the factors. 
    }
    \label{fig:study-outline}
\end{figure}

\subsection{Predictions of our cost-benefit framework}
Applying the cost-benefit framework to human-AI decision making, we see that the cost (in this case, cognitive effort) required to engage with the task depends on the difficulty of the underlying task and the difficulty of using the explanation. Given these two factors, we predict that:

\begin{quote}
    \textsc{Prediction 1:} As tasks increase in difficulty, there will be greater reductions in overreliance with explanations, compared to only getting predictions. 
    \\
    \textit{This is because explanations are relatively more ``useful,'' or have more utility, in harder tasks, since they greatly reduce the cognitive effort required to engage with the task.}
\end{quote}

\begin{quote}
    \textsc{Prediction 2:} As the effort to understand the explanation decreases, overreliance decreases.
    \\
    \textit{Intuitively, explanations that are easier to parse are thus easier to discern when they are correct or incorrect.}
\end{quote}

Applying the cost-benefit framework, we can also see that benefit required to engage with the task can depend on factors such as monetary reward (\ie in a crowdsourced task) or stakes (\ie in real-world medical decision-making). As such, we predict that:

\begin{quote}
    \textsc{Prediction 3:} As the monetary benefit of correctly completing the task increases, overreliance decreases.
    \\
    \textit{People are more inclined to expend effort into a task if there are higher stakes attached to it.}
\end{quote}

Finally, our theory predicts that an individual's \textit{subjective utility} of the AI is associated with the cognitive effort reduction given by the explanation. 
In the cost-benefit framework, people estimate the value of a strategy, called its subjective utility, by assessing the degree to which it maximizes benefits and minimizes costs. To understand the cost-benefit framework's relevance in Human-AI interaction, we need to address the connection between cost, benefit, and subjective utility in the Human-AI interaction domain. Therefore, we predict that: 

\begin{quote}
    \textsc{Prediction 4:} As tasks increase in difficulty, people will attach higher subjective utility to an AI's explanation.
    \\
    \textit{This is because, in harder tasks, the explanation greatly reduces the cost (therefore increasing the utility) of engaging with the task.}
\end{quote}

\begin{quote}
    \textsc{Prediction 5:} As the effort to understand the explanation decreases, people will attach higher subjective utility to an AI.
    \\
    \textit{This is because explanations that are easier to parse reduces the cost of (therefore increasing the utility) engaging with the task.}
\end{quote}

We summarize these factors in Figure~\ref{fig:study-outline}.

\section{Methods}
Our research aims to study the effects of explanations on human-AI decision-making. In the following section, we outline our experimental setup in which participants accomplish a task in collaboration with an AI system, with and without explanations, in different task difficulty settings. \edit{We situate our work in the context of human-AI decision making tasks. In such a task, a human is presented with some input and asked to make a decision, which may be an open ended answer or a multi-choice answer. To aid them with the task, the decision maker is partnered with an AI, which provides a prediction for the answer. In some conditions, the AI can also provide an explanation for their prediction. The decision maker is ultimately responsible for producing the final answer and has the option of utilizing or ignoring the AI's prediction and explanation.}

\begin{figure}[tb]
    \centering
    \includegraphics[width=\textwidth]{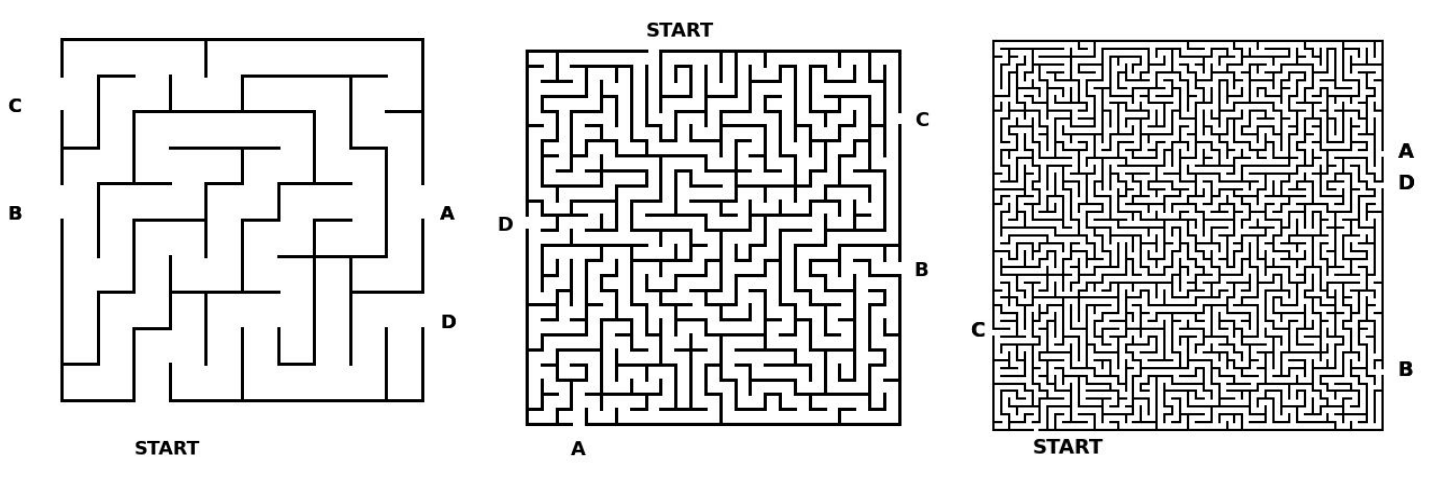}
    \caption{ Examples of the maze task that participants are asked to collaborate with the AI. \textbf{Left:} An example of our easy maze, which is a 10x10 grid. \textbf{Middle:} An example of our medium-difficulty maze, which is a 25x25 grid. \textbf{Right:} An example of our hard maze, which is a 50x50 grid.
    }
    \label{fig:task-difficulty}
\end{figure}

\subsection{Human-AI collaboration task}

For the purposes of our study, we needed a task that fulfills the following requirements:
\begin{enumerate}
    \item \textit{Multi-class classification}:
    In order to measure overreliance, we require that a random guessing strategy not produce the same answer as the AI very often: that is, we require a decision of higher cardinality than binary classification.
    \item \textit{Non-trivial for humans to complete alone}:
    We require a task that presupposes the need for an AI's help. The AI should be able to help on the task by giving predictions; the explanations should reduce the cost of engaging with the task.
    \item \textit{Lack of domain knowledge}:
    Since we are not testing our framework on a specific use-case, we needed a task that did not require specialized domain knowledge. 
    \item \textit{Enables different levels of task difficulty}:
    We require a task that can yield different levels of difficulty so that we can test their respective effects on overreliance.
    \item \textit{Enables multiple types of explanations}:
    We require a task that can yield different explanation modalities so that we can test their respective effects on overreliance.
\end{enumerate}
Prior XAI literature has used a variety of tasks, including binary classification~\cite{bansal2021does,bansal2019beyond,Lai_2019,bucina2020proxy} and question answering about images~\cite{chandrasekaran2017tango,chu2020visual} and text~\cite{bansal2021does,feng2019ai}. However, these tasks do not allow for the manipulations necessary for our experiments. Binary classification tasks, such as sentiment classification~\cite{bansal2021does}, would lead to $50\%$ overreliance by random guessing, so it would be more difficult to measure the effects of task conditions on the participant's unconscious decision to use an overreliance strategy.
Other tasks, such as answering questions about text~\cite{rajpurkar2016squad} or predicting the calorie count in images of food~\cite{buccinca2021trust}, are not amenable to controlling task difficulty (where ``easy'' is very little cognitive effort, and ``hard'' is very high cognitive effort).

\begin{figure}[tb]
    \centering
    \includegraphics[width=\textwidth]{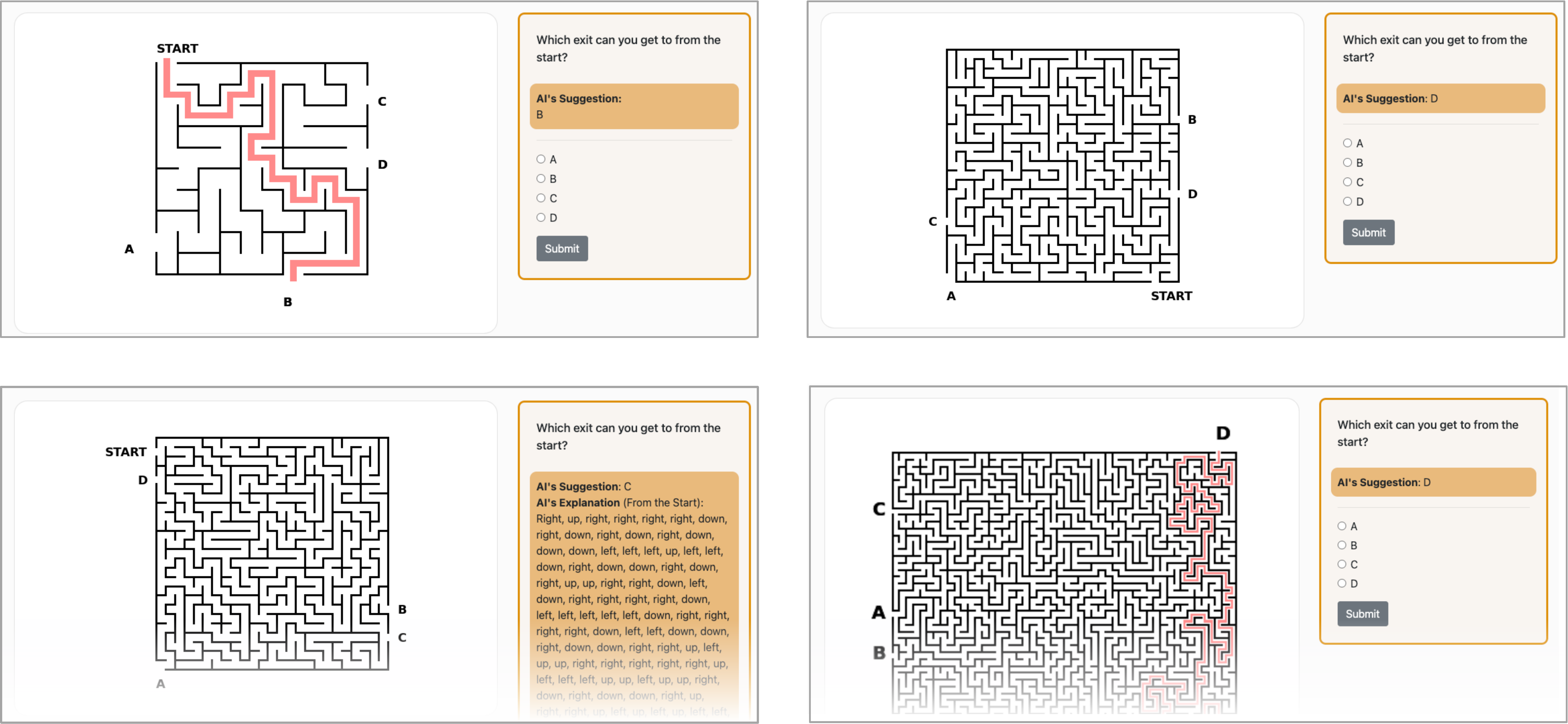}
    \caption{
    An example of a maze solving task in which the AI's suggestion is correct. We visualize three different task difficulty conditions and two different explanation conditions. \textbf{Top Left:} The \textit{easy task, highlight explanation} condition. The maze is $10 \times 10$ and features highlights of the path the AI suggests. The AI's suggestion is provided above the answer choices. \textbf{Top Right:} The \textit{medium task, prediction} condition. The maze is $25 \times 25$ and does not show
    explanation highlights.  \textbf{Bottom Left} The \textit{medium task, written explanation} condition. The maze is $25x25$ and features a written version of the path the AI suggests. \textbf{Bottom Right:} The \textit{hard task, highlight explanation} condition. The maze is $50 \times 50$ and features highlights of the path the AI suggests.}
    \label{fig:task-condition}
\end{figure}

In our paper, we use a visual search maze solving task, which meets all of our stated needs:
\begin{enumerate}
    \item \textit{Multi-class classification}:
    Participants are shown a maze with a start position with four possible exits. Participants are asked to determine which of the four possible exits is the true exit. With $4$ possible answers, the random chance of choosing the correct answer is $25\%$. When the AI is incorrect, overreliance with random guessing is also $25\%$.
    \item \textit{Non-trivial for humans to complete alone}:
    Mazes, especially large mazes, can be difficult to complete alone; so an AI could conceivably help by making a prediction of which exit is correct. Furthermore, engaging with the task is difficult as the participant still must solve the maze to check. Therefore, including (easy to understand) explanations reduces the cost of engaging with the task.
    \item \textit{Lack of domain knowledge}:
    Mazes do not require expertise or domain knowledge to complete.
    \item \textit{Enables different levels of task difficulty}:
    We can manipulate the difficulty of the task by changing the dimensions of the maze.
    \item \textit{Enables multiple types of explanations}:
    The maze task supports multiple explanation modalities, including highlight explanations, which are easy to check, and written explanations, which are more difficult to check. We visualize examples of these explanations in Figure~\ref{fig:task-condition}.
\end{enumerate}

\noindent\textit{Ecological Validity:}
Selecting maze solving trade\edit{s off} some ecological validity compared to other tasks, as few human-XAI tasks outside of the lab will include solving mazes, but a rich literature in cognitive psychology and cognitive science involves solving puzzles of this sort (e.g.,~\cite{anzai1979theory,lee2008psychological}). 

Even though AI are not often used to solve maze tasks in a collaborative setting, we argue that our maze task \edit{is comparable to }
% \edit{\sout{is similar to that of}} 
game playing settings, some of which, like Chess or Go, are often used in human-AI collaborative settings~\cite{chen2016evolution}. \edit{Though these situations are not identical in nature (e.g., maze tasks do not have the same societal importance as a game like Chess, which may affect peoples' behavior), we argue that predictions and explanations made by an AI in the maze task have a similar structure as those made by an AI for a game like Chess.} Our prediction in the maze task can be likened to the chess example of predicting the next move; our explanation in the maze task can be likened to the chess example of showing future roll-outs of the game (to justify its initial prediction). 
\\\\
\noindent\textit{Maze Selection:}
We programmatically generate mazes to be $10 \times 10$, $25 \times 25$, and $50 \times 50$  dimensions in height and width, corresponding to easy, medium, and hard tasks, respectively.
To ensure that the mazes are of an appropriate difficulty, we pre-tested the questions on crowdworkers. We ran a study in which six crowdworkers on the Prolific platform answered the questions with no help from the AI. Questions from the easy mazes that did not yield $100\%$ human accuracy were omitted (because we wanted the easy questions to truly be easy). And questions from the medium and hard mazes that did yield $100\%$ human accuracy were omitted (because we wanted the medium and hard questions to truly reflect these difficulties).

\subsection{AI Model}

\noindent\textit{Simulated AI:} 
To study overreliance in human-AI decision making, it is important that the AI accuracy is comparable to human accuracy~\cite{bansal2021does}. When humans and AIs perform roughly equally for a given task, there is no additional incentive for the human decision-maker to over- or underrely on the AI's predictions. In pilot studies, the human accuracy was $83.5\%$ across all difficulty conditions. Therefore, we control the number of incorrect AI predictions each participant sees such that AI accuracy is exactly $80\%$ for all studies (unless specified in the study design section of a particular study). For the questions the AI gets incorrect, about 1/3 of them come from mazes with lengthy path solutions, 1/3 from mazes with medium-length path solutions, and 1/3 from mazes with short path solutions. This ensures that participants in our study do not notice a bias in the type of mazes the AI gets incorrect.
\\\\
\noindent\textit{Simulated Explanations (And why they're considered explanations):} 
\edit{For our main studies}, we generate two types of explanations: highlight and writing explanations (Figure~\ref{fig:task-condition}). Although what constitutes an explanation is debated by philosophers~\cite{pitt1988theories}, we employ the following definition from ~\cite{explanations1980s} that characterizes explanations as describing how the ``explanandum'' (that which is being explained, or, in our case, the AI's predictions) is logical consequence of given conditions or premises. \edit{The explanations we created for this work, much like similar work in XAI, are local explanations in a decision-making context~\cite{miller2018explanation}. These explanations attempt to provide people with insights of \textit{how} the AI came to its decision. While our explanations give this decision-making information by indicating the path the AI took, other explanations, like SHAP~\cite{SHAP}, provide features that contributed most to the AI's outcome.}
\\\\
\noindent\textit{Highlight Explanations:} Our highlight explanations indicate the path between the start point and the predicted exist. Highlight explanations have commonly been used in prior XAI studies~\cite{bansal2021does,hase2020evaluating,feng2019ai,lai2020chicago,zhang2020effect,nguyen-2018-comparing}. We argue that our highlight explanations constitute as an explanation given by the definition from ~\cite{explanations1980s}. The conclusion (the exit to take to get out of a maze) follows from the premise (the steps taken to an exit in a maze). Since the highlight explanation describes visually the premise, this is an explanation.

Our two other, highlight based explanations, \textit{incomplete} and \textit{salient} show the highlights in a similar manner. However, in these cases, if the AI has crossed a wall (an illegal move in the maze setting) to reach its answer, the incomplete explanation does not show the path after the wall and the salient explanation highlights the path after the wall in blue. This is designed to make the AI's errors more obvious.
\\\\
\noindent \textit{Written Explanations:} For written explanation, we translated the ground-truth path into words (e.g. ``left, up, right, up...'').
Our written explanations are more analogous to logical written arguments explaining the process of arriving at the answer~\cite{bansal2021does}, and in our context, \textit{increase} the cost of engaging with the task (using the explanations) relative to the highlight explanation, as participants have to parse the writing and map it onto the maze. We argue that our written explanations also constitute a viable explanation, since it describes, in a written format, the premise (the steps taken to an exit in a maze) which logically leads to the conclusion (the exit to take to get out of the maze).
\\\\
\noindent \textit{Generation of explanations: }We generate ``accurate'' explanations when an AI's prediction is correct. When the AI is incorrect, our explanations indicate a path that goes through a maze wall. Evaluating with ``accurate'' explanations ensures that our results are not confounded by faulty or uninformative AI-generated explanations.

\subsection{General Study Design}
All of our five studies follow a similar template (unless otherwise noted in the particular study's procedure section). The number of trials and exact details are also noted in the procedure section of every study. 

All participants are onboarded, where they are asked demographic questions (age and gender). Then, they complete a training phase where they complete the mazes of their task difficulty condition alone. Then, they complete a training phase with the AI in their task difficulty and AI condition. In both training phases, they are given immediate feedback on whether their answers were correct or not. \edit{We chose to include this training phase as preliminary pilot participants displayed unstable behavior before familiarizing themselves with the AI. This allowed participants to become familiar with the task, the AI, and the AI's explanations, if applicable.} After both training phases, they are placed in a collaboration phase with the AI in their task difficulty and AI condition. They are not given feedback on whether their answers were correct or not. \edit{We removed feedback on answer correctness as we did not want their judgements of the AI's ability to change across the collaboration phase.}

Finally, we collect subjective self-report measures on how much participants trusted the AI, on how they interacted with the AI, and on their need for cognition (NFC) score~\cite{coelho2020nfc}, a stable personality trait measuring people's propensity towards completing cognitively difficult tasks. \edit{We measured this as it might be indicative of person-dependent differences in cognitive engagement in our task.} Unless otherwise noted, each self-report question is assessed using a 7-point Likert scale. The full set of self-report questions and NFC questions are provided in the appendix.
\\\\
\noindent\textit{Mitigating effects of trust: }
In light of trust's well-documented relation to reliance~\cite{lee2004trust,buccinca2021trust,bansal2021does,lai2019human,zhang2020effect,bussone2015role,yu2019trust}, we aim to neutralize any effects of trust in our task such that any differences in overreliance are attributable to our cost-benefit manipulations and not to individual differences in trust. To do so, we remove any task or interface components with known relations to trust. In particular, we abstain from anthropomorphizing the AI, always referring to the AI as ``the AI'', as Anthropomorphization has been linked to increased trust~\cite{hoff2015trust}. We always refer to AI predictions as ``suggestions''~\cite{park2019slow} to counteract perfect automation schemas~\cite{dzindolet2002perceived} and emphasize that this suggestion can be rejected. 
As participants complete a series of maze solving tasks, we do not provide immediate feedback since humans are particularly sensitive to seeing algorithms err~\cite{dietvorst2015algorithm}.
Lastly, we do not provide any information about the model's accuracy, as making this explicit also modulates trust~\cite{lai2019human,yu2019trust,yin2019understanding,zhang2020effect}.  
\subsection{Participant Selection}
All of our participants are recruited via Prolific (\href{https://www.prolific.co/}{prolific.co}), an online crowdsourcing platform. We only included participants that had at least $50$ submissions, were located in the United States, were native English speakers, had an approval rating of at least $95\%$ on Prolific, and that had not completed our task before. All participants received a payment of $\$4$ for $20$ minutes of their time with no bonuses, unless otherwise noted in a specific study's procedure section. We estimated the time required to complete the tasks through pilot experiments. 

\subsection{Measured Variables}
We collect the following objective performance measures for our hypotheses:
\begin{enumerate}
    \item \textit{Overreliance:} We define overreliance as the percentage of decisions when a human decision maker accepts an AI's incorrect prediction~\cite{buccinca2021trust}. For example, if a decision-maker agreed with the AI's prediction 80\% of the time when the AI was incorrect, then the overreliance is 80\%.
    \item \textit{Need for Cognition (NFC) Score\edit{~\cite{coelho2020nfc}}:} \edit{NFC is a stable personality trait measuring people's propensity towards completing cognitively difficult tasks.} We average the responses to the six NFC questions \edit{(e.g., ``I would prefer complex to simple problems.''), which are available in the appendix,} 
    % \edit{\sout{(with questions 3 and 4 on the scale reverse coded)}} 
    to compute a person's NFC score.
\end{enumerate}

\section{Study 1 -- Manipulating Costs via Task Difficulty}

\begin{figure}[t]
    \centering
    \includegraphics[width=0.75\textwidth]{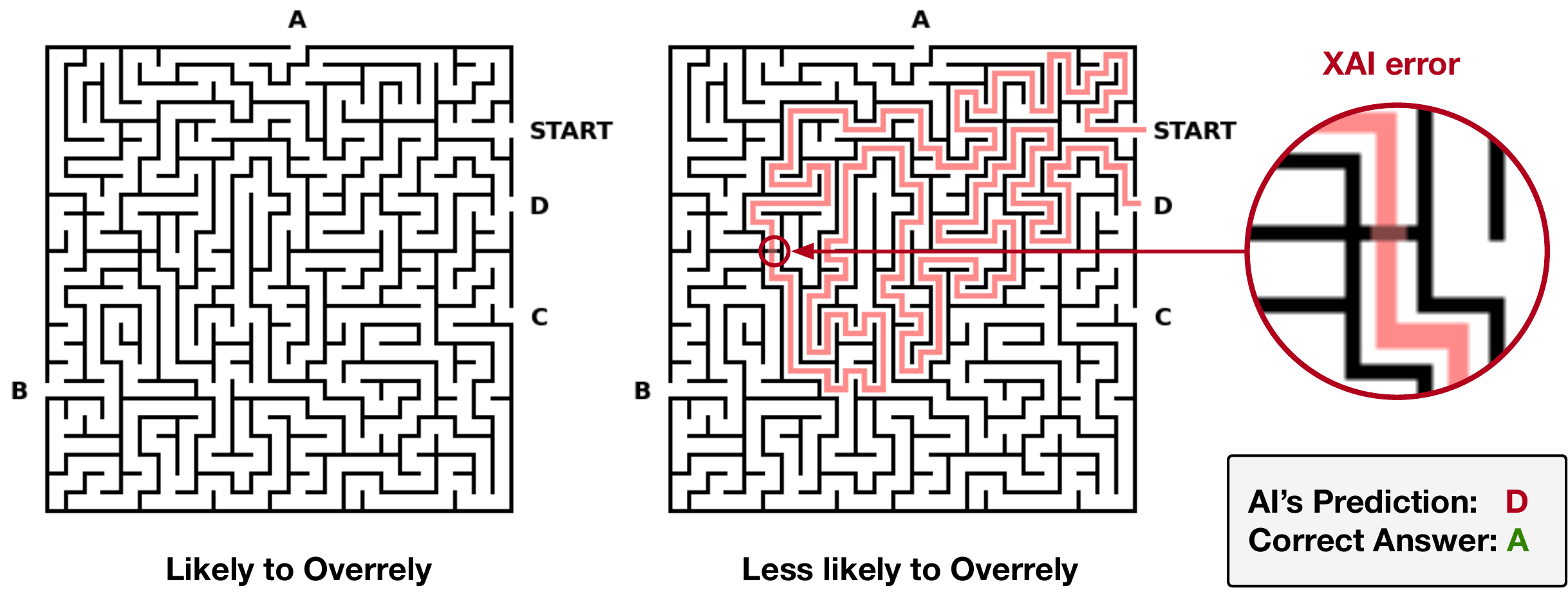}
    \caption{\textbf{Left:} The \textit{medium-difficulty task, prediction only} condition. The maze is $25 \times 25$ and features the prediction only. \textbf{Right:} The \textit{medium-difficulty task, explanation} condition. The maze is $25\times25$ and features explanations in the form of inline highlights. An example of when the AI gives an incorrect prediction and explanation.}
    \label{fig:incorrect-AI}
\end{figure}

In our first study, we explore \textsc{Prediction 1} by manipulating how difficult the task is to do in the presence or absence of explanations. The presence of an explanation (e.g., highlighting of relevant information) can decrease the cognitive effort required to engage with the task. However, explanations may not dramatically reduce the effort required to solve a task if the task is easy for a human to solve on their own. 
We pre-registered this study 
(\href{https://osf.io/vp749}{osf.io/vp749} and \href{https://osf.io/g6tjh}{osf.io/g6tjh}).
\subsection{Conditions}
We adopt a two-factor mixed between and within subjects study design in which each participant sees one of two \textit{AI conditions} in multiple task \textit{difficulty} conditions. Half of the participants see both easy and medium task difficulty conditions, and the other half of participants see only the hard condition, since the amount of time required to complete the task was approximately the same in these two study configurations.
The differences between conditions are depicted in Figure~\ref{fig:task-difficulty} and~\ref{fig:task-condition}.
\\\\
\noindent\textbf{Task difficulty conditions:} The task difficulty condition manipulates the maze difficulty. This manipulation modifies cognitive effort because it requires more cognitive effort to search for the correct path in a longer maze~\cite{mcclendon2001complexity}. Note that quality of questions, predictions, and explanations remain the same in both task difficulty conditions.
\begin{enumerate}[noitemsep,topsep=2pt]
    \item \textit{Easy:} In the easy condition, participants see a $10 \times 10$ maze.
    \item \textit{Medium:} In the medium condition, participants see a $25\times25$ maze.
    \item \textit{Hard:} In the hard condition, participants see a $50 \times 50$ maze.
\end{enumerate}

\noindent\textbf{AI conditions:} The AI condition manipulates the presence of explanations for an AI's prediction.
\begin{enumerate}[noitemsep,topsep=2pt]
    \item \textit{Prediction:} In the prediction condition, participants are only provided with the AI's prediction. The AI's predicted answer is displayed directly below the question.
    \item \textit{Explanation:} In the explanation condition, participants are provided with a highlight explanation in addition to the AI's prediction. Highlights are displayed directly on the maze.
\end{enumerate}

\subsection{Hypotheses}
Our first prediction expects that explanations will reduce overreliance when they reduce the cost of engaging with the task; similarly, explanations will not reduce overreliance when the reduction of cost in engaging with the task is small (e.g. in the easy task). Additionally, we include a hypothesis to test if there are individual differences in propensity to engage with the task (\textsc{H1e}).

Therefore \textsc{Study 1} has 5 hypotheses:
\begin{quote}
    \textsc{Hypothesis 1a:} In the \textcolor{AIColor}{prediction only} case, overreliance is \textbf{lower} in the \textcolor{TaskColor}{easy task} than in the \textcolor{TaskColor}{medium-difficulty task}.
    \end{quote}
\begin{quote}
    \textsc{Hypothesis 1b:} In the \textcolor{TaskColor}{easy task}, overreliance is \textbf{the same} in the \textcolor{AIColor}{prediction only} case and   \textcolor{AIColor}{prediction with explanation} case.
\end{quote}
\begin{quote}
    \textsc{Hypothesis 1c:} In the \textcolor{TaskColor}{medium-difficulty task}, overreliance is \textbf{lower} in the  \textcolor{AIColor}{prediction with an explanation} case than in the \textcolor{AIColor}{prediction only} case.
\end{quote}
\begin{quote}
    \textsc{Hypothesis 1d:} In the \textcolor{TaskColor}{hard task}, overreliance is \textbf{lower} in the  \textcolor{AIColor}{prediction with an explanation} case than in the \textcolor{AIColor}{prediction only} case.
\end{quote}
\begin{quote}
    \textsc{Hypothesis 1e:} In the \textcolor{TaskColor}{hard task}, overreliance is affected by an \textbf{interaction effect} between participants' Need for Cognition (NFC) scores and the \textcolor{AIColor}{AI Condition}.
\end{quote}
\begin{figure}[t]
    \centering
    \subfigure[][Overreliance]{
        \includegraphics[width=0.42\textwidth]{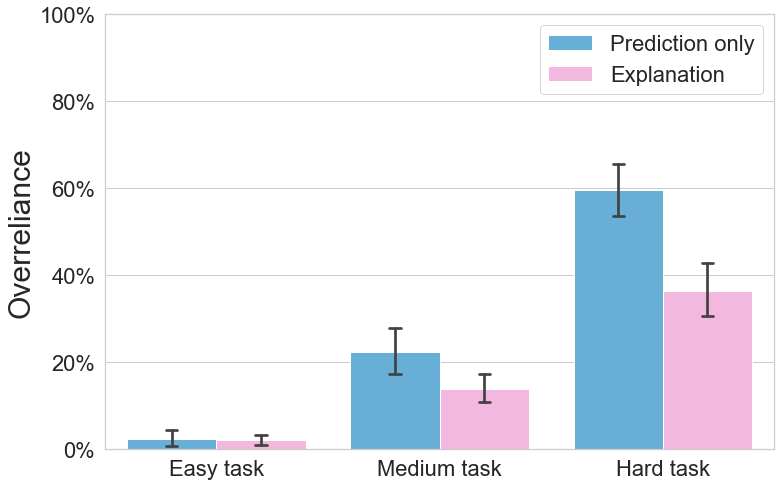} 
    }
    \subfigure[][Accuracy]{
        \includegraphics[width=0.42\textwidth]{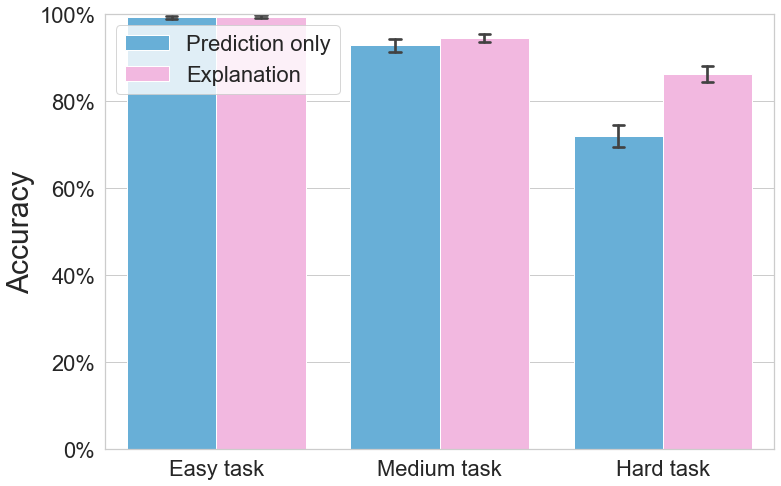}
    }
        \caption{\textbf{(a)} Overreliance levels in study 1. In the easy and medium-difficulty tasks, we find no differences between prediction and explanation conditions. We find that explanations reduce overreliance in the hard task condition. \textbf{(b)} Accuracy levels in study 1. In exploratory analysis, we find that explanations increase decision-making accuracy in the hard task condition. These findings together suggest that the gains provided by explanations are more likely to be found when tasks are hard for people to do.}
    % \end{minipage}
    \label{fig:study1-plot}
\end{figure}

\subsection{Procedure}
For participants in the easy and medium condition, during the initial human-only phase, participants are asked to do two easy mazes and two medium mazes alone. During the initial AI training phase, participants are prompted to complete ten questions with an AI aid to experience using it, where half are easy and half are medium. The AI agent always offers a prediction for each question; in the explanation condition, it also highlights the path in the maze. We ensure that the AI errs on two of the ten training questions (one in easy and one in medium) and randomize the order in which participants encounter these ten questions. 

After the training phase, participants are asked to complete thirty test questions in collaboration with the same AI agent. Exactly six of the thirty questions have incorrect AI predictions and their positions are randomized. 

For participants in the hard condition, given how long it takes to complete the task, we decrease the total number of trials. Therefore, we cut the number of mazes in half in each phase of the study and only showed participants the hard condition. This, however, does not affect how hard people perceived the task to be, since (1) people did not have access to how many trials they had left, and (2) qualitative data shows people found the hard task to be more difficult than the other two tasks.

\subsection{Participants}
We recruited a total of $N=340$ participants ($N=170$ in prediction only, $N=170$ in explanation, half of which are in the easy and medium condition, and the other half of which are in the hard condition). Our sample size comes from a power analysis done on pilot data.

\subsection{Results}
We analyze our results using a Bayesian\footnote{\edit{We adopt Bayesian statistics for its advantages in being able to hypothesize null effects~\cite{Navarrete17, gelman1995bayesian}. Bayesian statistics has also been proposed as a better alternative to frequentist statistics for HCI research~\cite{kay2016researcher}. For more information on our analysis, please see the appendix section.}} linear mixed effects model (see Table~\ref{tab:study1},~\ref{tab:study1_part2}). We visualize our results in Figure ~\ref{fig:study1-plot}. We adopt the convention of saying that a comparison is notable if the 95\% credible interval of the posterior distribution excludes 0 and the estimate of the mean is positive or negative (depending on the direction of the hypothesis). \textit{Note}: We provide the model specification in the appendix.

% We preregistered that this hypothesis would not be significant, so our hypothesis is confirmed.

\subsection{Summary} 
We find support for \textsc{H1a}, which provides empirical evidence for the following claim: as tasks increase in difficulty, overreliance on AI's predictions increases. This can be explained in the following way through our cost-benefit framework: an increase in task difficulty increases the cost of following the strategy to engage with the task, as it generally takes more cognitive effort for people to either (a) ensure that the AI has gotten the answer correct when provided with only predictions or (b) complete the task alone. However, the strategy to rely on the AI does not incur the same costs, as people are not engaging with how hard the task is. Since the strategy to rely on the AI inherently results in overreliance, overreliance will increase, as seen here.
\\\\
\begin{table}[t]
\caption{\edit{Results from study 1. We find that overreliance increases as tasks get harder; overreliance does not change with explanations in easy tasks; and overreliance decreases with explanations in hard tasks. This supports our hypothesis that explanations are more likely to provide assistance, and therefore reduce overreliance, when tasks are hard. This fits into our cost-benefit framework, as explanations in hard tasks sufficiently reduce the cost of doing a task properly.}}
\label{tab:study1}
\begin{center}
\resizebox{\linewidth}{!}{
    \begin{tabular}{lccccccc} 
     \toprule
     Parameter & Mean & Lower 95\% CI & Upper 95\% CI & Notable & Hypothesis & Confirmed?\\
     \midrule
     Easy Pred - Medium Pred & -3.967 & -5.480 & -2.53 & * & \textsc{H1A} &  \textbf{yes} \\ 
     Easy Pred - Easy XAI & 0.164 & -1.678 & 2.17 & & \textsc{H1B} &  \textbf{yes} \\ 
     Medium Pred - Medium XAI & 0.916 & -0.305 & 2.08 & & \textsc{H1C}  & no\\ 
     \midrule
     Hard Pred - Hard XAI & 1.74 & 0.889 & 2.76 & * & \textsc{H1D} & \textbf{yes}\\
     \bottomrule
    \end{tabular}
}
\end{center}
\end{table}

\begin{table}[t]
\caption{\edit{Results from study 1. We do not find support for our hypothesis that there is an interaction effect of AI Condition and a participants' Need for Cognition (NFC) score. We hypothesized that people who are more likely to engage in cognitively demanding tasks would have little differences in overreliance levels regardless of whether AI explanations were provided or not. However, we did not find support for this, which we hypothesize could be due to the fact that, in the hard task, even participants with high propensities to engage in effortful thinking are likely to overrely. An initial exploratory analysis of an interaction between AI Condition and NFC in the medium task condition provides some evidence to support this claim.}}
\label{tab:study1_part2}
\begin{center}
\resizebox{\columnwidth}{!}{
    \begin{tabular}{lccccccc } 
     \toprule
     Parameter & Mean & Lower 95\% CI & Upper 95\% CI & Notable & Hypothesis & Confirmed?\\
     \midrule
     AI Condition:NFC & 0.35 & -0.76 & 1.49 & & \textsc{H1E} &  no
     \\
     \bottomrule
    \end{tabular}
}
\end{center}
\end{table}
\noindent
We find support for \textsc{H1b}, which provides empirical evidence for the following claim: in easy tasks, predictions and explanations will provide similar levels of overreliance. This can be explained in the following way through our cost-benefit framework: since, in the easy task, explanations do not make engaging with the task any easier (\ie they do not provide any reduction in costs), the utility of the two strategies do not change. Therefore, overreliance levels do not change. 
\\\\
\noindent
We do not find support for \textsc{H1c}; therefore not providing empirical evidence for the following claim: in medium-difficulty tasks, there is more overreliance in the prediction condition than in the explanation condition. This may be in due part to the fact that this condition is not hard enough to see the reductions in cost that we expect to see with explanations. 
\\\\
\noindent
We find support for \textsc{H1d}, which provides empirical evidence for the following claim: in hard tasks, there is more overreliance in the prediction condition than in the explanation condition. This can be explained in the following way through our cost-benefit framework: in hard tasks, explanations provide extra assistance to completing the task over predictions only, greatly reducing the cost of engaging with the task. Since the costs have been greatly reduced and are relative to the costs of what would otherwise be incurred (e.g. by completing the task alone or just using an AI's prediction), the explanation looks far more ``useful'' as a means of reducing effort. 
Therefore, overreliance levels reduce in this case since people are more likely to use the strategy to engage in the task.
\\\\
\noindent
We do not find support for \textsc{H1e}; therefore, not providing empirical evidence for the following claim: there is an interaction effect between participants' Need for Cognition (NFC) scores and the type of explanation modality (highlights of lackthereof) when measuring overreliance. This could be because the hard task is too difficult to demonstrate differences in behavior across peoples' NFC, since most people are likely to overrely on the AI's prediction anyway. In an exploratory analysis, we find a interaction between AI condition and Need for Cognition (NFC) in the medium task. We also find that, as NFC scores increase, the difference between overreliance in prediction and explanation diminishes. The medium task could be a sufficiently difficult enough task to demonstrate differences in behavior across NFC.
\\\\
\noindent
Additionally, we ran an exploratory analysis to measure the effects of explanations on human-AI decision-making accuracy. Our exploratory analysis uncovers an instance where explanations increase human-AI accuracy. In the hard task, explanations increase the accuracy compared to the prediction only condition (see Figure~\ref{fig:study1-plot}b). This finding stands in contrast against prior work~\cite{bansal2021does,buccinca2021trust}, which has failed to find differences between the conditions.
\\\\
\noindent
In summary, explanations do not produce an observable reduction in overreliance in easy or medium tasks, but do reduce overreliance in the hard task. This result demonstrates that overreliance can in fact be mitigated by explanations---and that this occurs precisely when tasks are substantially complex and effortful that reviewing the explanation yields a substantial cognitive effort benefit.
\section{Study 2 -- Manipulating Costs via Increasing Explanation Difficulty}
\begin{figure}[t]
    \centering
    \includegraphics[width=\textwidth]{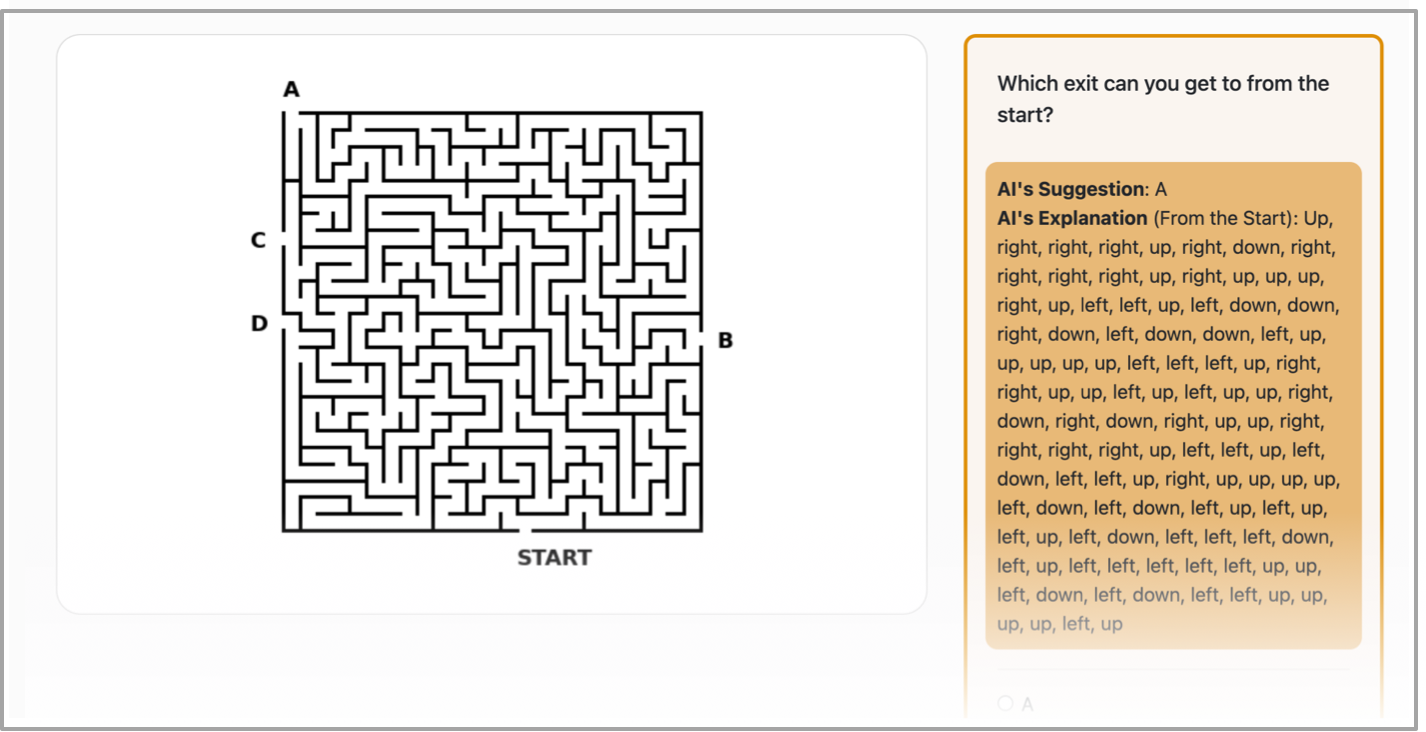}
    \caption{
    The \textit{medium task, written explanation} condition. The maze is $25 \times 25$ and features a written version of the path the AI suggests. In the written explanation, each word corresponds to one step in the maze. We find that these written explanations do not reduce overreliance, compared to not receiving an explanation.}
    \label{fig:task-condition-2}
\end{figure}

In our second study, we explore \textsc{Prediction 2} by manipulating how difficult the explanation is to parse. We hypothesize that explanations that are easier to understand will have lower levels of overreliance. We pre-registered this study 
(\href{https://osf.io/4dbqp}{osf.io/4dbqp}) 
.
\subsection{Conditions}
We adopt a two-factor mixed between and within subjects study design in which each participant sees one of two \textit{explanation difficulty} conditions in multiple task \textit{difficulty} conditions. Half of the participants see both easy and medium task difficulty conditions, and the other half of participants see only the hard condition, since the amount of time required to complete the task was approximately the same in these two study configurations.

Since we wanted to manipulate the difficulty of the explanation in order to measure its effects on overreliance, we created an easy-to-understand explanation and a hard-to-understand explanation. We did so by creating visually easy-to-parse highlight explanations overlaid on the maze. These easy explanations are contrasted with written explanations that incur more cognitive effort, since participants need to follow along these written explanations that are on the side of the maze (see Figure~\ref{fig:task-condition-2}).
\\

\noindent\textbf{Explanation conditions:} The explanation condition manipulates the difficulty of the explanations provided to people.
\begin{enumerate}[noitemsep,topsep=2pt]
    \item \textit{Highlight explanation:} Participants see the predictions as well as explanations, which are displayed by directly highlighting the proposed path to the AI's chosen exit. 
    \item \textit{Written explanation:} Participants see the predictions as well as explanations, which are displayed by writing the path alongside the maze (e.g. ``left, left, up, right, down...'') that the AI takes from the start.
\end{enumerate}

\subsection{Hypothesis}
Our second set of hypotheses further addresses the connection between overreliance and the cognitive effort required to engage with a task by manipulating explanation costs. 
Specifically, in our second study, we adjust the cost of engaging with the task by manipulating explanation difficulty. Explanations can decrease the cognitive effort of engaging with the task, insofar as the content of the explanation makes doing the task sufficiently easier. \textsc{Prediction 2} expects that explanations that are easier to parse have less overreliance.
Therefore \textsc{Study 2} has 2 hypotheses:
\begin{quote}
    \textsc{Hypothesis 2a:} In the \textcolor{TaskColor}{medium-difficulty task}, overreliance is \textbf{lower} in the \textcolor{AIColor}{highlight explanation} case than in the \textcolor{AIColor}{written explanation} case.
\end{quote}
\begin{quote}
    \textsc{Hypothesis 2b:} In the \textcolor{TaskColor}{hard task}, overreliance is \textbf{lower} in the \textcolor{AIColor}{highlight explanation} case than in the \textcolor{AIColor}{written explanation} case.
\end{quote}
\subsection{Procedure}
We adopt the same study procedure for Study 1.

\subsection{Participants}
We recruited an additional $N=170$ participants from Profilic, using the same exclusion criteria as in Study 1. All were placed in the written explanation condition. Half were placed in the easy and medium task difficulty condition; half were placed in the hard task difficulty condition. The highlight explanation condition has $170$ participants, from the data collected in Study 1. This gives us a total of $N=340$ participants in Study 2. Our sample size comes from a power analysis done on pilot data. 
\begin{figure}[t]
    \centering
    \includegraphics[width=0.6\textwidth]{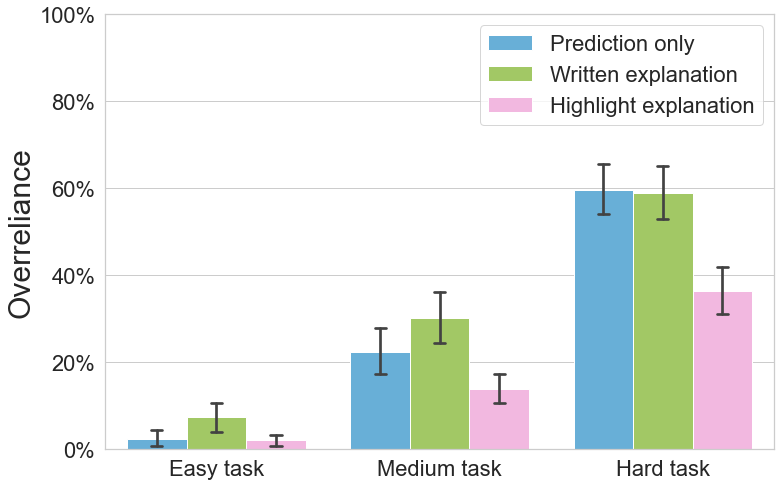} % first figure itself
    \caption{Overreliance levels in study 2. We find that overreliance is greater in the written explanations condition than in the highlight explanation condition. \edit{This supports our hypothesis that explanations that are easier to parse will generally yield lower levels of overreliance.}}
    \label{fig:study2}
\end{figure}

\subsection{Results}
We analyze our results using two Bayesian linear mixed effects models (for \textsc{H2a} and \textsc{H2b}, respectively; see Table~\ref{tab:study2}). We visualize our results in Figure~\ref{fig:study2}. We adopt the convention of saying that a comparison is notable if the 95\% credible interval of the posterior distribution excludes 0 and the estimate of the mean is positive or negative (depending on the direction of the hypothesis). \textit{Note}: We provide the model specification in the appendix.

\begin{table}[t]
\caption{\edit{Results from study 2. We find support for our hypotheses that there are lower levels of overreliance when provided an explanation in the form of highlights (easy to parse), as opposed to in the form of written language (hard to parse).}}
\label{tab:study2}
\begin{center}
\resizebox{\columnwidth}{!}{
    \begin{tabular}{lcccccc}
     \toprule
     Parameter & Mean & Lower 95\% CI & Upper 95\% CI & Notable & Hypothesis & Confirmed? \\
     \midrule
     Highlight - Written in Medium& -1.63 & -2.78 & -0.559 & * & \textsc{H2a}& \textbf{yes} \\ 
     \midrule
    Highlight - Written in Hard & -1.632 & -2.570 & -0.694 & * & \textsc{H2b} & \textbf{yes} \\ 
     \bottomrule
    \end{tabular}
}
\end{center}
\end{table}

\subsection{Summary}
We find support for \textsc{H2a}, which provides empirical evidence for the following claim: as explanations increase in difficulty to understand, overreliance on the AI increases. This can be explained in the following way through our cost-benefit framework: an increase in explanation difficulty increases the cost (via cognitive effort or time) to engage with the task. This increase in cost makes the strategy to engage with the task seem less appealing or ``useful.'' As people engage less with this strategy, they might follow the other strategy: rely on the AI. As people are more likely to choose the strategy to rely on the AI in this case, they will have higher levels of overreliance.
\\\\
\noindent
We find support for \textsc{H2b}, which provides empirical evidence for the following claim: as explanations increase in difficulty to understand, overreliance on the AI increases. This can be explained in the same way as \textsc{H2a} can.
\\\\
\noindent
Additionally, we ran exploratory analyses which showed, across all task difficulty (easy, medium, and hard) conditions, there was no difference between the prediction only case and the hard-to-understand, written explanation case. This suggests that these hard-to-understand explanations do not act as an additional signal for trust (which would increase overreliance).
\\\\
\noindent
In summary, Study 2 demonstrates that overreliance is responsive not just to the cognitive effort of completing the task, but also to the effort to understand an explanation.
\section{Study 3 -- Manipulating Costs via Decreasing Explanation Difficulty}

In our third study, we explore how decreasing the difficulty of understanding explanations can reduce overreliance. In accordance with the results of study 2, we postulate that explanations with lower levels of difficulty (to understand) will have lower levels of overreliance. This experiment tries to push the boundaries of displaying \textit{extremely obvious} explanations to see if there is a discernible floor effect to overreliance (\eg will there always be \textit{some} overreliance, or can we get it to zero?) This experiment is exploratory and therefore does not have pre-registration.
\subsection{Conditions}
We adopt a between-subjects experiment, wherein participants are placed in one of five explanation conditions. 
Participants are placed in the hard task difficulty condition, since we wanted to see how much we could reduce overreliance (and the hard task has the highest levels of overreliance in the prediction only case).
\\\\
\noindent\textbf{AI conditions:} The AI condition manipulates the presence of and difficulty of the explanations provided to people.
\begin{enumerate}[noitemsep,topsep=2pt]
    \item \textit{Prediction:} Participants are only provided with the AI's prediction.
    \item \textit{Highlight explanation:} Participants see the predictions as well as explanations, which are displayed by directly highlighting the proposed path to the AI's chosen exit. 
    \item \textit{Written explanation:} Participants see the predictions as well as explanations, which are displayed by writing the path alongside the maze (e.g. ``left, left, up, right, down...'') that the AI takes from the start.
    \item \textit{Incomplete explanation:} Participants see the predictions as well as a highlighted path to the AI's chosen exit. If the AI crosses a wall in the maze, the rest of the path is not shown.
    \item \textit{Salient explanation:} Participants see the predictions as well as a highlighted path to the AI's chosen exit. If the AI crosses a wall in the maze, the rest of the path is highlighted in blue.
\end{enumerate}
\begin{figure}[tb]
    \centering
    \includegraphics[width=\textwidth]{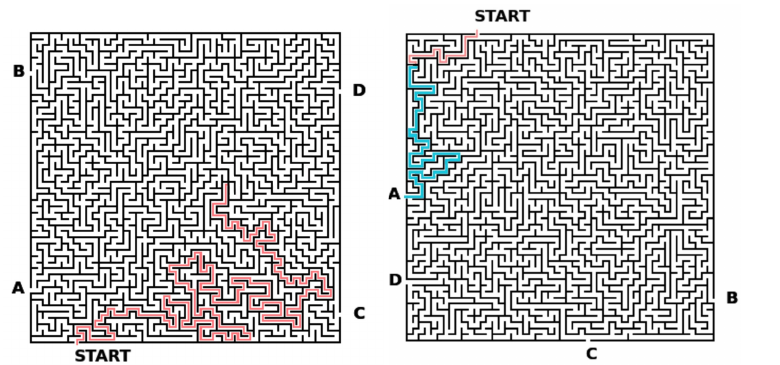}
    \caption{
    An example in which the model is incorrect in (a) the incomplete explanation, where the the explanation does not continue after the AI goes over a wall, and (b) salient explanation, where the explanation is highlighted in blue when the AI goes over a wall. \edit{We designed these explanations to push the boundaries of explanations' saliency. In study 3, we validated that, the more salient an explanation makes an AI's mistake, the less overreliance there is.}
    }
    \label{fig:study3}
\end{figure}
\subsection{Hypothesis}
We did not have pre-registered hypotheses for two reasons: (1) this is an exploratory study to find if there is a floor effect and (2) our hypotheses relating explanation difficulty and overreliance have been confirmed in \textsc{study 2}. However, we were guided by the following question: \textit{how much can we reduce overreliance by having explanations that make it easy to spot mistakes?}

\subsection{Procedure}
We adopt the same study procedure for Study 1 in the hard task.
\subsection{Participants}
We recruited an additional $N=31$ participants from Profilic, using the same exclusion criteria as in Study 1. All were placed in the hard condition. $16$ were in the incomplete explanation condition; $15$ were in the salient explanation condition. (Our sample size comes from a power analysis done on pilot data.) There were $N=85$ in each of the prediction only, highlight explanation, and written explanation conditions from study 1 and 2. In total, there were $N = 286$ participants in this study. 

\begin{figure}[tb]
    \centering
    \includegraphics[width=0.6\textwidth]{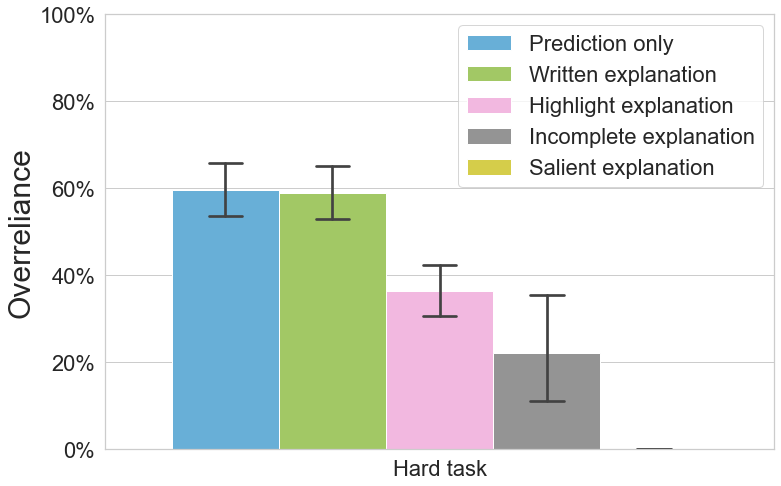} % first figure itself
    \caption{Overreliance levels in our exploratory study, study 3. \edit{We find that (1) more salient explanations reduce overreliance over less salient explanations and (2) there is no ``floor'' effect on overreliance; that is, we observe that people will not overrely if they \textit{know} the AI is wrong.}}
    \label{fig:study3-results}
\end{figure}

\subsection{Results}
\edit{We analyze our results using a Bayesian linear mixed effects model (see Table~\ref{tab:study3}). We visualize our results in Figure~\ref{fig:study3-results}. Because we didn't pre-register any hypotheses for this study, we report the means and credible intervals for all the pairwise comparisons here, but don't indicate whether the differences are credible. \textit{Note}: We provide the model specification in the appendix.}

\begin{table}[t]
\caption{\edit{Results from our exploratory study, study 3. We find that, the more salient an explanation is, the less overreliance there is. Additionally, we did not observe a ``floor'' effect on overreliance; that is, we observe that people will not overrely if they \textit{know} the AI is wrong.}}
\label{tab:study3}
\begin{center}
\resizebox{\columnwidth}{!}{\begin{tabular}{ lcccccc }
 \toprule
 Comparison & Mean & Lower 95\% CI & Upper 95\% CI \\
 \midrule
 Prediction - Incomplete Explanation & 3.336 & 1.309 & 5.36 \\ 
 Written Explanation - Incomplete Explanation & 3.205 & 1.236 & 5.38 \\
 Highlight Explanation - Incomplete Explanation & 1.479 & -0.490 & 3.60 \\
  Prediction - Salient Explanation & 41.698 & 5.055 & 153.76 \\
 Written Explanation - Salient Explanation & 41.570 & 5.316 & 153.95 \\ 
 Highlight Explanation - Salient Explanation & 39.816 & 3.843 & 152.03 \\
 Incomplete Explanation - Salient Explanation & 38.232 & 1.971 & 150.90 \\
 \bottomrule
\end{tabular}
}
\end{center}
\end{table}

\subsection{Summary}
Our exploratory analysis finds that the salient explanation condition has less overreliance than all other conditions (prediction, highlight explanation, written explanation, and incomplete explanation). We also find that the incomplete explanation condition has less overreliance than the prediction only and written explanation conditions, more overreliance than the salient explanation condition and finds no difference in overreliance in the highlight explanation condition. \edit{Complementary to Study 2, Study 3 provides further support for our hypotheses.}
\\\\
\noindent
Furthermore, we do not find a ``floor'' effect, \edit{\ie an effect in which we cannot reduce overreliance to $0\%$. The salient explanation condition, our most most obvious explanation condition, has an average overreliance rate of $0\%$.} Although there is more work to be done whether this is the case with more participants, these initial findings suggest that people are unlikely to agree with an AI if they \textit{know} it is incorrect.
\section{Study 4 -- Manipulating Benefit via Monetary Bonus}
In our fourth study, we explore \textsc{Prediction 3} to validate the ``benefit'' aspect of our cost-benefit framework. In this study, we aim to show that increasing the benefit of doing the task properly (\ie getting the answer correct) reduces overreliance. We pre-registered this study (\href{https://osf.io/hgz2x}{https://osf.io/hgz2x}).
\subsection{Conditions}
We adopt a two-factor mixed between and within-subjects experiment in which each participant sees both \textit{bonus conditions} and sees one of two \textit{AI conditions}. All participants are placed in the hard task difficulty condition. Half of the participants see the prediction only condition; half of the participants see the highlight explanation condition.
\\\\
\noindent\textbf{AI conditions:} The AI condition manipulates the presence of explanations for an AI's prediction.
\begin{enumerate}[noitemsep,topsep=2pt]
    \item \textit{Prediction:} Participants are only provided with the AI's prediction. The AI's predicted answer is displayed directly below the question.
    \item \textit{Explanation:} Participants are provided with a highlight explanation in addition to the AI's prediction. Highlights are displayed directly on the maze.
\end{enumerate}

\noindent\textbf{Bonus conditions:} The bonus condition manipulates how much bonus each maze completed is worth.
\begin{enumerate}[noitemsep,topsep=2pt]
    \item \textit{Low Bonus:} Participants are given \$0.01 for each maze done correctly.
    \item \textit{High Bonus:} Participants are given \$0.50 for each maze done correctly.
\end{enumerate}
\begin{figure}[tb]
    \centering
    \includegraphics[width=0.6\textwidth]{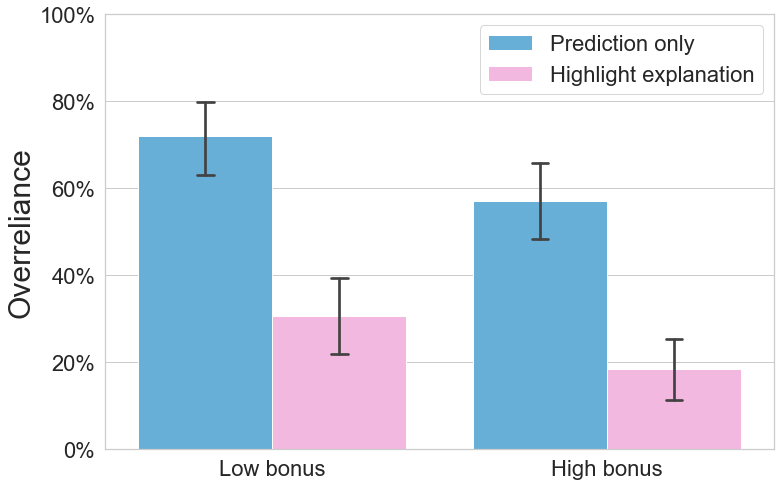} 
    \caption{Overreliance levels in study 4. \edit{We hypothesize and validate that, the more bonus provided for completing the task properly, the less overreliance there is.}}
    \label{fig:study4}
\end{figure}
\subsection{Hypothesis}
Our third set of hypotheses addresses the connection between overreliance and benefit of completing the task.
Specifically, in our fourth study, we adjust the benefit of each strategy by controlling the amount of bonus given. This influences how desirable a strategy such as rely on the AI looks to a participant, given that they will not get the bonus if the AI is wrong and they overrely.
Therefore \textsc{study 4} has two hypotheses:
\begin{quote}
    \textsc{Hypothesis 3a:}
     Overreliance is \textbf{lower} in the \textcolor{BonusColor}{high bonus} case than in the \textcolor{BonusColor}{low bonus} case.
\end{quote}
\begin{quote}
    \textsc{Hypothesis 3b:}
    Overreliance is \textbf{lower} in the \textcolor{AIColor}{highlight explanation} case than in the \textcolor{AIColor}{prediction} case.
    \textit{Same hypothesis as \textsc{H1d}; we retest this to ensure that our study changes in bonus and amount of questions do not affect our previous finding.}
\end{quote}
\subsection{Procedure}
We follow the same study procedure as study 1 in the hard task,
\edit{with two exceptions. First,} we adopted a blocked, within-subjects study design in which participants are randomly placed in one bonus condition for the first half of the study and the other condition for the second half\footnote{The pre-registration does not specify a blocked study design.}. \edit{We chose this study design because prior work~\cite{tversky1974judgment} shows that people calculate their judgements of monetary gains and losses relative to a certain reference point, which was indicative to us that we ought to have within-subjects comparisons of monetary value.} \edit{Second,} we add one question where the AI is incorrect to the collaboration phase as to have an even amount of questions with the low- and high- bonus.
\subsection{Participants}
We 
% \edit{\sout{will}} 
recruit\edit{ed} $N =114$ participants for our study, half of whom (57 participants) are in the prediction condition, the other half (57 participants) are in the explanation condition. Our sample size comes from a power analysis done on pilot data.
\subsection{Results}
We analyze our results using two Bayesian linear mixed effects models (for \textsc{H3a} and \textsc{H3b}, respectively; see Table~\ref{tab:study4}). We visualize our results in Figure~\ref{fig:study4}. We adopt the convention of saying that a comparison is notable if the 95\% credible interval of the posterior distribution excludes 0 and the estimate of the mean is positive or negative (depending on the direction of the hypothesis). \textit{Note}: We provide the model specification in the appendix.

\begin{table}[t]
\caption{\edit{Results from study 4. We find that overreliance decreases with more bonus provided for completing the task properly. This affirms the ``benefit'' element of our cost-benefit framework, showing that changes in reward or incentive structures affect peoples' behavior. Additionally, we reproduce a finding from study 1 that there are lower levels of overreliance when given a highlight explanation in a hard task, even when bonuses are added.}}
\label{tab:study4}
\begin{center}
\resizebox{\columnwidth}{!}{
    \begin{tabular}{lcccccc} 
     \toprule
     Comparison & Mean & Lower 95\% CI & Upper 95\% CI & Notable & Hypothesis & Confirmed? \\
     \midrule
     Low Bonus - High Bonus & 0.93 & 0.414 & 1.46 & * & \textsc{h3a} & \textbf{yes} \\ 
     \midrule
       Prediction - Highlight XAI & 2.47 & 1.72 & 3.36 & * & \textsc{h3b} & \textbf{yes} \\ 
     \bottomrule
    \end{tabular}
}
\end{center}
\end{table}

\subsection{Summary}
We find support for $\textsc{H3a}$, which provides empirical evidence for the following claim: when the benefit (in this case, via monetary bonus) of doing a task properly increases, overreliance decreases. This can be explained by our cost-benefit framework in that increasing the benefit of completing the task properly makes the utility of the strategy to engage with the task higher than the strategy to rely on the AI, since the latter strategy is likely to lead to more mistakes. Since the strategy to rely on the AI leads to the most overreliance and this is the strategy of which the utility is the lowest when benefits (of completing the task properly) increase, we can then say that the overreliance is decreased because of this utility comparison.
\\\\
\noindent
We find support for \textsc{H3b}, which provides empirical evidence for the following claim: in hard tasks, highlight explanations reduce overreliance compared to only receiving the AI's prediction. This replicates the finding in \textsc{H1d}.
\\\\
\noindent
Taken together, Study 3 demonstrates that overreliance is responsive not just to costs, but also to the benefit accrued for expending these costs. \edit{These findings also have implications on the designs of crowd-sourced studies, for example, as findings may differ when providing a bonus or not in a study.}
\section{Study 5 -- Measuring the utility of XAI methods}
In our fifth and final study, we measure the effect of task difficulty and explanation difficulty on the subjective utility participants assign to the AI. We explore \textsc{Prediction 4} and \textsc{Prediction 5} to validate our final two claims: (1) people attach higher utility to an AI's explanation when tasks are harder and (2) people attach higher utility to an AI's explanation when it is easier to understand. This study aims to make explicit the choice people make between strategies (\eg whether to rely on the AI or not). This final study validates our framework by showing that peoples' utilities of an AI are predicted by it. We pre-registered this study 
(\href{https://osf.io/cskvb}{https://osf.io/cskvb}).
\begin{figure}[tb]
    \centering
    \includegraphics[width=\textwidth]{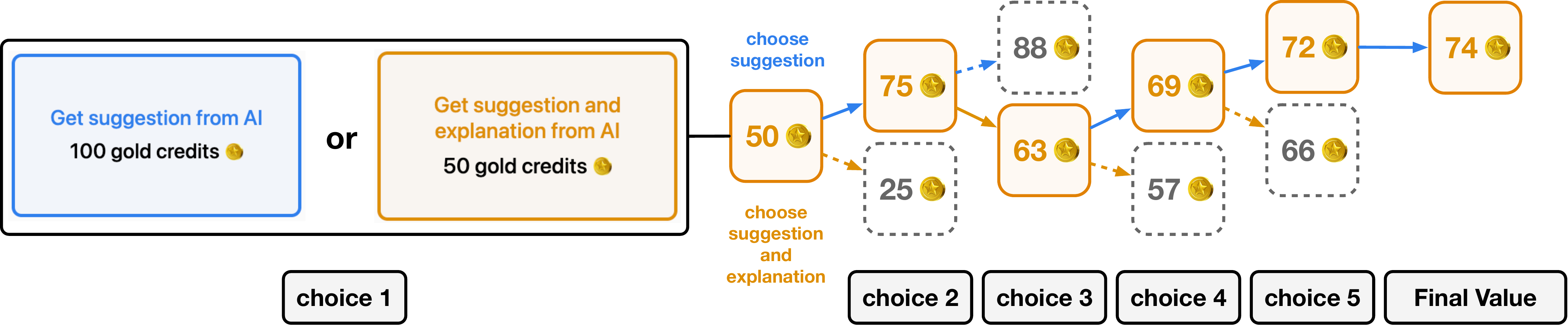}
    \caption{
    An overview of the cognitive effort discounting (COG-ED) task in Study 5. Participants are asked to choose between completing with the AI's suggestion for 100 gold credits or completing with the AI's suggestion and explanation for a dynamically shifting price.
    }
    \label{fig:cog-ed}
\end{figure}
\subsection{Cognitive Effort Discounting (COG-ED)}
To quantify participant's subjective utility of adopting various cognitive strategies when collaborating with an AI, we adapt the Cognitive Effort Discounting (COG-ED) paradigm~\cite{westbrook2013subjective} to human-AI collaboration (Figure \ref{fig:cog-ed}). During the standard COG-ED protocol, as originally designed~\cite{westbrook2013subjective}, participants are presented with a series of choices between a high-effort task for a fixed monetary reward and a low-effort task for a lower monetary reward. The COG-ED paradigm takes in this series of decisions and outputs a number representing the participant's perceived value of completing the low effort task.

COG-ED reaches this output number by adjusting the reward of the low effort tasks depending on the task participants choose (Figure~\ref{fig:cog-ed}). Consider an example in which a participant makes the first choice between completing hard task for a $\$1$ reward or an easier task for a $\$0.50$ reward. If the participant chooses the harder task, then the reward for the easier task increases for the next choice. If the participant chooses the easier task, then the reward for the easy task decreases for the next choice. The specific formula is detailed below. This process continues iteratively. This process is equivalent to a binary search that converges after a logarithmic number of choices. At this converged point, the subjective utility of both tasks are assumed to be equal, allowing the difference in reward to quantify the subjective utility of the low-effort task relative to the high-effort task. 

In two extreme conditions, if the participant always chooses the harder task, the reward for the easier task will increase and leave a small difference in reward between the harder and easier task. This small difference indicates that participant does not see the low-effort task as reducing the effort enough to be worth the lower reward. In contrast, if the participant always chooses the easier task, the reward for the easier task will decrease and leave a large difference in reward between the harder and easier task. This large difference indicates that the participant sees the low-effort task as reducing the effort enough to be worth receiving lower pay. Thus, the COG-ED paradigm indicates the person's valuation of the reduced effort in the low-effort task. 

In behavioral economic terms, the COG-ED paradigm is an example of a choice experiment, a common method for inferring subjective utilities~\cite{westbrook2015cognitive}. The assumption underlying such \textit{choice experiments} is that the indifference point, \ie the point at which a decision-maker freely chooses between two conditions at roughly equal rates, implies that the decision-maker associates both conditions with equal utilities~\cite{westbrook2015cognitive}. 
In human-computer interaction, similar choice experiments have previously been employed to assess the utility of user interfaces~\cite{toomim2011utility}.

\subsection{Conditions}
We modify existing COG-ED experiments for human-AI teams by asking participants to choose between completing the task alone (high-effort) versus completing the task with the AI (low-effort). Participants are placed in one of two within-subject scenarios. 
\\\\
\noindent\textbf{Within-Subjects Scenarios:} 
\begin{enumerate}[noitemsep,topsep=2pt]
    \item \textit{Comparison between AI's utility in different task difficulty conditions.} Participants work with two AI agents-- one of which can generate highlight explanations in the medium task, the other of which can generate highlight explanations in the easy task. 
    \\
    \textit{We presume that these results can also extrapolate to comparisons between easy and hard, as well as medium and hard. This is because the predicted effect size between easy and medium is smaller than the other comparisons.}
    \item \textit{Comparison between AI's utility in different explanation conditions.} Participants participate with two AI agents -- one of which can generate highlight explanations in the medium task, the other of which can generate written explanations in the medium task.
    \\
    \textit{We test these in the medium task since (1) the easy task was unlikely to find any differences in behavior and (2) we presume any results can also extrapolate to the hard task, since in prior studies we found that effect sizes of manipulations were much smaller in the medium task than the hard task.}
\end{enumerate}

\subsection{Hypothesis}
\textsc{Prediction 4} states that people will attach higher subjective utility to an AI's explanation in harder tasks. This can be adapted to our study with the following hypothesis:
\begin{quote}
    \textsc{Hypothesis 4a:}
    The subjective utility of the AI is \textbf{lower} in the \textcolor{TaskColor}{easy task} than in the \textcolor{TaskColor}{medium-difficulty task}.
\end{quote}
\textsc{Prediction 5} states that people will attach higher subjective utility to an AI's explanation when it is easier to understand. This can be adapted to our study with the following hypothesis:
\begin{quote}
    \textsc{Hypothesis 5a:} The subjective utility of the AI is \textbf{lower} in the \textcolor{AIColor}{written explanation} case than in the \textcolor{AIColor}{highlight explanation} case.
\end{quote}

\subsection{Procedure}

Similar to the first two studies, this study is divided up in an AI training and a testing phase (there is no human-only training) in each within-subjects condition. We made only one change to the study as described in the preregistration: instead of completing two questions without the AI at the beginning, the participant completes two questions with the AI (prediction only) in the beginning. Then, participants complete five questions with the AI (\edit{
% \sout{prediction and }
}
explanation) in the AI training phase and five questions with the AI (\edit{
% \sout{prediction and }
}explanation) in the testing phase. 

Unlike previous studies, this study has a COG-ED phase after the testing phase.
During the COG-ED experiments, participants are asked to complete $5$ questions, during which, they are asked to choose between completing the task in a high-effort prediction-only condition for a fixed financial reward ($R_{max}$ credits, where $R_{max} = 100$ in our experiments) or completing the task in a low-effort explanation condition for a smaller reward (initially $R_1$ credits, where $R_1 = 50$ in our experiments). The rewards are initialized such that $R_1 < R_{max}$. The high-effort condition, in this case, is opting to complete the task alone. The low-effort condition, in this case, is opting to complete the task with the AI. Each time a participant chooses the low-effort option, the reward for the low-effort option shifts to adjust for the participant's preferences. Specifically, assuming the reward for the low-effort condition is $R_t$, the new reward is dynamically updated such that:
\begin{align}
    R_{t+1} = \begin{cases}
    \frac{R_{t} + R_{max}}{2},& \text{if high-effort is chosen}\\
    \frac{R_{t} + R_{t-1}}{2},& \text{otherwise}
    \end{cases}
\end{align}
where $R_{0} = 0$. The monetary reward for choosing the high-effort prediction-only condition does not change.

\edit{After the participants have completed the five questions in the COG-ED phase, the monetary reward $R_{6}$ is the final value on which the participants have converged. $R_{max} - R_{6}$ represents how much reward a participant is willing to forgo in order to do an easier task. This quantifies the \textit{subjective utility} of the AI in the easier task.}

\begin{figure}[tb]
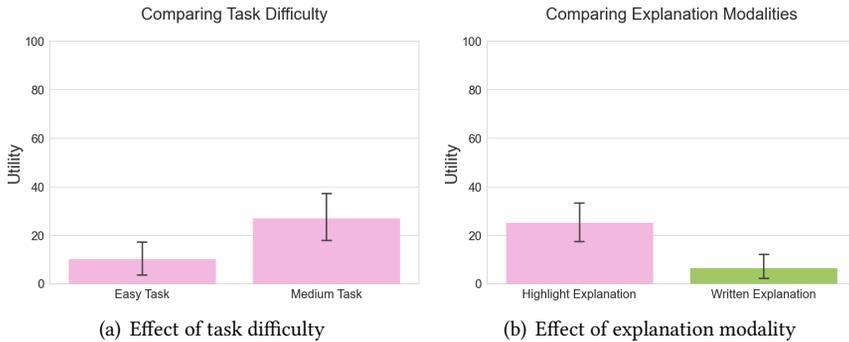

    \centering
    \subfigure[][Effect of task difficulty]{                                  
        \includegraphics[width=0.4\textwidth]{study_results/study5-diff.pdf}
    }
    \subfigure[][Effect of explanation modality]{
        \includegraphics[width=0.4\textwidth]{study_results/study5-xai.pdf}
    }
    
    \caption{\textbf{(a)} Utility levels in study 5.  We find that highlight explanations have a higher subjective utility in the medium task than in the easy task.  \textbf{(b)} Utility levels in study 5. We find that highlight explanations have a higher subjective utility than written explanations. \edit{Taken with the findings of studies 1 and 2, this study shows that people tend to attach higher utility to AI Conditions that generally decrease overreliance. This study validates that people are indeed computing a ``cost-benefit'' utility and that it is highly predictive of general trends in overreliance.}}
    \label{fig:study5}
\end{figure}
\subsection{Participants}
We recruited $N=76$ participants from Profilic, half of which are placed in condition $1$, half of which are placed in condition $2$, using the same exclusion criteria as in Study 1. Our sample size comes from a power analysis done on pilot data.

Participants receive a base payment of $\$4$ for $20$ minutes of their time, the same as Study 1 and Study 2. However, participants are also given opportunities for bonuses in this study when they answer a question correctly. To equalize participant perception of monetary reward and mitigate income effects, we use a credit system, where participants earn ``credits'' instead of dollar amounts. Participants are informed during the study that $100$ credits are equivalent to $\$0.10$ in the medium task and $100$ credits are equivalent to $\$0.05$ in the easy task (for participants who see the easy task). 
\subsection{Results}
We analyze our results for \textsc{H4} and \textsc{H5} using two Bayesian linear mixed effects models (see Table~\ref{tab:study5}). We visualize these results in Figure~\ref{fig:study5}. We adopt the convention of saying that a comparison is notable if the 95\% credible interval of the posterior distribution excludes 0 and the estimate of the mean is positive or negative (depending on the direction of the hypothesis). \textit{Note}: We provide the model specification in the appendix.
\begin{table}[t]
\caption{\edit{Results from study 5. We find that people place a higher utility level on the AI's explanations when the task is harder. We also find that people place a higher utility level on the AI's explanations when it is easier to parse. These results, taken together with the results of study 1 and 2, provide evidence for our cost-benefit framework in overreliance, as people place higher utility on the AI Conditions that generally yield lower levels of overreliance.}}
\label{tab:study5}
\begin{center}
\resizebox{\columnwidth}{!}{\begin{tabular}{ ccccccc }
 \toprule
 Comparison & Mean & Lower 95\% CI & Upper 95\% CI & Notable & Hypothesis & Confirmed? \\
 \midrule
  Medium task - Easy task & 16.96 & 5.22 & 28.7 & * & \textsc{H4a} & \textbf{yes} \\ 
 \midrule
 Highlight XAI - Written XAI & 18.61 & 9.51 & 27.79 & * & \textsc{H5a} & \textbf{yes} \\ 
 \bottomrule
\end{tabular}
}
\end{center}
\end{table}
\subsection{Summary}

We find support for \textsc{H4a}, which provides empirical evidence for the following claim: people find AI's explanations to have higher utility in harder tasks. This fits into our cost-benefit framework: explanations are used with the strategy to engage with the task. In easier conditions, this strategy is about as much work as doing the task alone, since the AI provides little additional help; however, as the task gets harder, this strategy allows people to (1) be more likely to be correct with (2) less cost (via time and cognitive effort). This strategy therefore greatly reduces the cost needed to complete the task properly (compared to doing it alone) in hard tasks.
\\\\
\noindent
We find support for \textsc{H5a}, which provides empirical evidence for the following claim: people find AI's explanations to have higher utility when they are easier to understand. This fits into our cost-benefit framework: explanations are used with the strategy to engage with the task. Anyone who engages with this strategy would want to also minimize the cost to do so. Since explanations that are harder to understand have higher costs (in the form of cognitive effort and time) than those that are easier to understand, they have lower subjective utility.
\\\\
\noindent
This study demonstrates that people are indeed making a strategic decision to overrly on the AI. Our findings match with the outcomes of the cost-benefit framework, which predicts overreliance on the AI.
\section{Discussion}
Taken together, our results identify scenarios in which explanations reduce overreliance. In line with prior work examining cost-benefit analyses in the allocation of mental effort~\cite{kool2018mental}, we provide evidence that humans engage in cost-benefit analyses when choosing \textit{how} to collaborate with an AI when making decisions.

\subsection{Summary of results}
Our five studies demonstrate the validity of our cost-benefit framework. We find that explanations reduce overreliance in more difficult tasks. We also find that the \textit{type} of explanation matters: explanations that make the AI's mistake more obvious reduce overreliance more. This supports our ``cost'' element of the framework: people will engage with the task when elements (e.g. the explanation by an AI) reduce cognitive costs (e.g. in hard tasks or if the explanations are very easy to verify the AI's predictions).
We also find that overreliance is reduced when people see a benefit to not overrelying: when people are given a larger monetary benefit to complete the task, they overrely less. Finally, we validate our claim that people strategically choose when to engage with the AI. We do this by calculating the subjective utility and find that, overwhelmingly, people attach higher utility to explanations in harder tasks and that are easier to understand. We summarize our primary findings on overreliance (studies 1-4) in Table \ref{tab:summary}. 
\\\\

\begin{table}[t]
\caption{Summary of primary findings. We find that there is an overall increase in overreliance as tasks are more difficult; that in easier tasks, overreliance decreases with explanations; that in hard tasks, overreliance decreases with explanations; that given explanations that are easier to parse or that make the AIs mistake more salient, overreliance decreases; and that when the benefit to getting the answer correct is higher,  overreliance decreases.}
\label{tab:summary}
\resizebox{\columnwidth}{!}{\begin{tabular}{llllccccc}
    \toprule
    Hypothesis & Task Difficulty & AI Condition & Monetary Bonus & Overreliance\\
  \midrule
  \textsc{H1a} & Easy $\rightarrow$ Medium & Prediction & &	$\uparrow$
  \\
  \textsc{H1b} & Easy & Prediction $\rightarrow$ Highlight XAI & & same
  \\
  \textsc{H1c} & Medium & Prediction $\rightarrow$ Highlight XAI & & same
  \\
  \textsc{H1d} \& \textsc{H3b} & Hard & Prediction $\rightarrow$ Highlight XAI & & $\downarrow$
  \\
  \textsc{H2a} & Medium &  Written XAI $\rightarrow$ Highlight XAI& & $\downarrow$
  \\
  \textsc{H2b} & Hard & Written XAI $\rightarrow$ Highlight XAI & & $\downarrow$
  \\
  Exploratory & Hard & XAI becomes more salient & & $\downarrow$
  \\
  \textsc{H3a} & Hard & Prediction \& Highlight XAI & Low $\rightarrow$ High & $\downarrow$
\end{tabular}
}
\end{table}

\subsection{Analysis of prior work}
Conflicting expectations (of explanations reducing overreliance) and empirical results (of explanations not reducing overreliance) largely inspired this paper. As such, in this section, we aim to address how the findings of prior work fit into our framework. \citeauthor{bansal2021does}~\cite{bansal2021does} find that explanations do not decrease overreliance; however, we postulate that the task (sentiment analysis) could have been too easy to complete alone and that the explanation modality (logical explanations) was too complicated to understand when participants were placed in a higher-effort LSAT task. \citeauthor{buccinca2021trust}~\cite{buccinca2021trust} find that cognitive forcing functions decrease overreliance compared to simple explanations. We postulate that this could be due to cognitive forcing functions increasing the difficulty of doing the task. ~\cite{gonzalez2020human} find that some explanations do help in error detection. The study's task was in open-domain question-answering, which is a more difficult, open-ended task, where explanations can reduce costs of engaging with the task.

\subsection{Implications}
Our results find instances in which explanations reduce overreliance over a prediction only baseline. We identify some of the metaphorical knobs to be turned to modulate levels of overreliance: task difficulty, explanation difficulty, and monetary reward. These, of course, were proxies for other elements of human-AI interaction; for example, monetary reward aimed to target increasing benefits, which could also be targeted by increasing the stakes of a task. In the following section, we discuss in more detail how designers can manipulate the task to reduce overreliance.
\\\\
\noindent\textit{Manipulating costs:}
Here, we discuss some design details that can influence overreliance by manipulating the associated costs (via cognitive effort, time, or other factors) of the task:
\begin{enumerate}
    \item \textit{Cognitive forcing functions:} As~\cite{buccinca2021trust} found, these cognitive forcing functions decrease overreliance. We postulate that these forcing functions increase the cost of following the strategy to rely on the AI.
    \item \textit{Applying easy-to-understand explanations to hard tasks:} Our work finds that explanations are especially beneficial when tasks are hard. Tasks that are already difficult may be good places to apply explanations.
    \item \textit{Artificially manipulating tasks to be harder:} As in our experiments, designers can also manipulate the relative costs of engaging with the task by increasing the baseline task difficulty. This outcome may arise naturally as AI performance improves and AI systems tackle more challenging tasks. However, in many settings, increasing the difficulty of the task may not be feasible.
    \item \textit{Creating easier-to-understand explanations:} Our work finds that, if explanations are not easy-to-understand, then they do not provide any assistance over prediction only baselines. Designs should include explanations that are easy to map to the AI's predictions for verification. However, generating human-interpretable explanations continues to be a challenge for XAI~\cite{humanInterpretableXAI2019}.
\end{enumerate}
\noindent\textit{Manipulating Benefits: }
Here, we discuss some design details that can influence overreliance by manipulating the associated benefits (via reward, stakes, or other factors) of the task:
\begin{enumerate}
    \item \textit{Monetary reward:} Although this is not true of all tasks, some tasks, like crowd-sourcing, could benefit from reward systems where people are bonused for their joint work with AIs. \edit{Since we found an impact of rewards on overreliance, there may also be a difference in the findings of work that bonuses crowd-workers versus work that does not. }
    \item \textit{Stakes:} The benefit of the task may also include professional, personal, or other benefits in the form of stakes. High-stakes domains, for example, may be a case where the benefits of doing tasks properly is really high, thereby discouraging people from relying on the AI.
    \item \textit{Enjoyability:} Tasks that are inherently enjoyable also have high benefit attached to them, as people naturally want to engage with the task~\cite{enjoyment2009}. Though not possible for all tasks, increasing the enjoyability of a task may increase peoples' propensity for engaging with the task.
    \item \textit{AI design:} Although increasing benefits of the following may result in increases in overreliance, designers are also able to manipulate (1) perceived model accuracy~\cite{yin2019understanding}, (2) AI anthropromorphization~\cite{kulms2019more}, and (3) the competence associated with metaphors describing the system~\cite{khadpe2020conceptual}. For example, anthropromorphizing the AI may lead to an increase sense of trust in the AI's accuracy, which may lead people to falsely assume that such an AI has a greater benefit (via overall accuracy) than it actually does.
\end{enumerate}

\subsection{Limitations and Future Work}
\noindent\textit{Task: }
As previously mentioned, our task was in a very controlled, low-stakes, maze-solving environment. As such, we do not know how much of our work will extend to the following settings (and others):
\begin{enumerate}
    \item \textit{High-stakes environments:} \edit{Our study focused on a low-stakes maze solving task. Future work should explore our theory on high-stakes tasks such as recidivism prediction~\cite{explanationshelpful2021}, where the consequences of a wrong decision are much higher. High stakes tasks may have different levels of costs and benefits, in which case our findings may not generalize. Such an experiment is out of scope for the current paper given the complexity that might arise due to Prospect theory~\cite{tversky1992advances}. Prospect theory explains that people value gains and losses differently, placing more weight on perceived gains versus perceived losses. We hypothesize that the perception of loss by making an incorrect decision in high-stakes tasks will modify the costs associated with the task. This in turn might affect overreliance independent of task difficulty and explanation conditions.}
    \item \textit{Generative tasks:} Generative tasks often lack a ``correct,'' ground truth answer because of their open-ended nature. Future work remains to see how much the results of this work would translate to more open-ended tasks.
    \item \textit{Non-game-playing tasks:} As noted by a few participants, our task was, at least to some, inherently ``enjoyable.'' As enjoyability may affect how much people want to engage with tasks, future work remains on how this work translates to tasks less gamified. 
    \item \textit{Uncertain environments:} Some decision-making tasks, such as toxicity prediction, lack a ``correct'' answer~\cite{gordon2021disagreement}. In this case, we do not know how our framework in understanding when explanations reduce overreliance would translate, since there is inherent uncertainty in the decision being made.
    \item \edit{\textit{Prediction tasks with incomplete explanations}: Many common applications, including common image classification tasks and natural language inference tasks, utilize highlight explanations that identify patches of images or phrases in sentences that were primarily responsible for a prediction. These types of highlight explanations provide incomplete information about a model's underlying behavior, unlike the procedural highlight explanations we used in the maze solving tasks. Future work should explore how well our theory translates to such prediction tasks.}
\end{enumerate}

\noindent\textit{Explanations: }
Our explanation modalities were primarily limited to those in the form of highlights and written explanations. The following are some crucial limitations to the explanations we chose:
\begin{enumerate}
    \item \textit{Explanation quality:} In our work, we specifically chose ``perfect'' explanations so that explanation quality would not affect our results. However, in practice, AIs do not provide these ideal explanations~\cite{brittleXAI}. \edit{Because study two shows that explanation understandability affects levels of overreliance, we hypothesize, but do not empirically validate, that lower explanation quality affects overreliance similarly. However, future work remains to assess the impact of explanation quality on our results, since there may be factors such as a change in trust that may arise when evaluating imperfect explanations.}
    \item \textit{Explanation format:} In our work, our explanations specifically helped users understand the AI's reasoning process to the end of the maze. However, some explanations, such as Shapley Additive Explanations~\cite{SHAP, nohara2019explanation} or contrastive example-based explanations~\cite{van2021evaluating}, only provide users a rough understanding to the AI's biases and output. These do not necessarily tell users about the veracity of the AI's prediction.
\end{enumerate}
\noindent\textit{Users:}
Our studies were primarily done with crowd-sourced workers. 
As mentioned in the \textit{Task} section, our study may not be reflective of high-stakes target domains of interest (such as medicine, law, risk assessment, etc.). The following outlines limitations and future work on users studied:
\begin{enumerate}
    \item \textit{Different practitioners:} While prior work has established ample evidence of cognitive biases in high-stakes domain~\cite{lambe2016dual,croskerry2009clinical,croskerry2009universal,lebovic2013policy,guthrie2007blinking,vasiljevic2013reasoning}, practitioners in these domains may have vastly different needs and values, yielding different assessments of costs and benefits. In light of recent calls for sociotechnical approaches to XAI~\cite{ehsan2020human,ehsan2021expanding,jacobs2021designing}, in which the human decision-maker's values and social context are centered when designing explanations, we emphasize the need for co-designing decision support tools with relevant stakeholders and evaluating these tools in real contexts. For instance, a recent study found that clinicians simply did not have time to engage with feature-based explanations for individual patient predictions and instead requested evidence that the tool had been clinically validated~\cite{jacobs2021designing}. In this example, the time pressures doctors face when treating patients inflated the costs of engaging with an explanation. \edit{As such, future work is necessary to test our framework on different practitioners and how different levels of expertise may affect overreliance.}
    \item \textit{Long-term users:} Our crowd-sourced study lacks long-term user engagement,. Future work remains to study how long-term users engage with these systems \edit{and how overreliance might change over time.}
    \item \textit{Need for Cognition:} Our study found mixed results whether a person's propensity for effortful thinking influenced their overreliance. As such, future work is necessary to tease apart the individual factors that contribute to this phenomena.
    
\end{enumerate}

\noindent\textit{Interface Design: }
People moderate their perceptions of the costs and benefits of using an AI using a wide variety of sources. Through pilot studies, we found that small changes in the interface affected the results: including a progress bar, the total number of questions, naming the AI, providing feedback on what answers were correct, and explicitly stating the model's accuracy. We chose the most ``neutral'' setting with each of these design choices and limited the number of questions asked for each user to avoid fatigue. Future work can explore how changes in these interface considerations affect behavior.
\\\\
\noindent\textit{Trust: }
This paper does not address the causes and effects of AI explanations on trust. 
Previous work has suggested that trust increases someone's likelihood of using a machine~\cite{zhang2020effect,yu2019trust}. However, XAI studies found conflicting results about the effect of explanations on trust. 
Some studies found that explanations increase trust~\cite{bucina2020proxy,bansal2021does}, some found no significant effect~\cite{zhang2020effect,chu2020visual,kaur2020interpreting}, while others found varying effects under different conditions~\cite{kulesza2013too,kunkel2019trust}.
Explanations intend to improve performance in part by helping a user calibrate their trust according to a  more accurate mental model of the AI's ability and accuracy~\cite{bansal2021does}. The effect of a mental model of AI accuracy is unclear, as some studies found that any accuracy score increases trust~\cite{lai2019human}, others found high accuracy scores increased trust and low accuracy scores decreased trust~\cite{yu2019trust,zhang2020effect}, while another found that these effects are mediated by an individual's innate likelihood of trusting automation~\cite{pop2015individual}.

It also remains unclear whether explanation-induced trust is caused by explanations increasing understanding of the underlying system or explanations serving as a blind signal for trust and reliance. A recent study found that placebic, uninformative explanations increased trust in a recommender system nearly as much as real explanations~\cite{eiband2019impact}, mirroring work in social psychology demonstrating adding a redundant explanation to a request (``May I use the xerox machine, \textit{because I have to make copies}?'') increases compliance~\cite{langer1978mindlessness}.

Finally, when researching the effects of trust on behavior, trust is difficult to measure accurately. Some studies use reliance as a proxy measure of trust~\cite{bansal2021does,lai2019human,zhang2020effect,chu2020visual}, while others subjectively measure an individual's trust in the system and innate disposition to trust~\cite{buccinca2021trust,bucina2020proxy,kaur2020interpreting,bussone2015role,kunkel2019trust,pop2015individual}. It will also be necessary to address how individuals in different contexts and occupations build trust in the systems they use. Although measuring and modulating trust was outside the scope of this paper, we see a rich opportunity for future work further investigating the effects of AI explanations on trust and the effects of trust on a cost-benefit analysis and resulting behavior.

\section{Conclusion}
Prior work in human-AI decision-making found that explanations do not empirically reduce overreliance. Inspired by cost-benefit analysis in behavioral economics, we outline a framework that aims to identify when explanations reduce overreliance. Our studies validate this framework. We find that when we increase the cost of doing the task alone or with just the AI's predictions (via increasing task difficulty), people utilize explanations, which leads to a reduction in overreliance, compared to a prediction only baseline. We also find that when we increase the cost of engaging with the task (via decreasing the understandability of an explanation), people are less likely to utilize the explanation, which leads to an increase in overreliance. We validate the benefit component and find that, as the benefit of completing the task properly increases (via monetary bonus), overreliance decreases. Through our studies using the Cognitive Effort Discounting paradigm, we quantify the utility of different explanations in different settings. We find that people explicitly forego monetary rewards for an AI in a harder task or for an AI that gives more understandable explanations. We observe that people assign high utility to the conditions that had the biggest reductions in overreliance (compared to the prediction only baseline) in prior experiments. Our experiments suggest that overreliance is not merely an immutable inevitability of cognition but at least, in part, a strategic choice.

\begin{acks}
We thank Snap Inc. for their funding of this project. We also thank the anonymous reviewers, Jenn Wortman Vaughan, Michelle Lam, and Joon Sung Park for feedback on this draft. 
\end{acks}

\bibliographystyle{ACM-Reference-Format}
\bibliography{references.bib}

\section{Appendix}
\subsection{Self-report Questions}
6-Question Trust Survey:
\begin{enumerate}
  \item I believe the AI is a competent performer. (on a 7 point Likert scale)
  \item I trust the AI. (on a 7 point Likert scale)
  \item I have confidence in the advice given by the AI. (on a 7 point Likert scale)
  \item I can depend on the AI. (on a 7 point Likert scale)
  \item I can rely on the AI to behave in consistent ways. (on a 7 point Likert scale)
  \item I can rely on the AI to do its best every time I take its advice. (on a 7 point Likert scale)
 \end{enumerate}
 
 \noindent Task Survey:
 \begin{enumerate}
     \item I found this task interesting. (on a 7 point Likert scale)
  \item I found this task difficult without the AI. (on a 7 point Likert scale)
  \item I found this task difficult even with the AI. (on a 7 point Likert scale)
  \item I would prefer to complete this task with the AI's suggestions than to complete it by myself. (on a 7 point Likert scale)
  \item Please select the option below which best represents how you used the AI. (4 options, listed below)
      \begin{enumerate} 
          \item I did not use the AI and completed the task by myself.
          \item I first completed the task myself and then verified my response with the AI's.
          \item I first looked at the AI's suggestion and then verified it was correct.
          \item I always chose the AI's suggestion. 
      \end{enumerate}
  \item Approximately, how accurate do you think the AI is? Please indicate using the slider below. (on a 100\% slider)
  \item In the box below, please describe how you used the AI when its suggestions were given to you. (free form answer box)
  \item In the box below, please describe how you chose between using the AI and doing the task yourself. (free form answer box)
\end{enumerate}

\noindent Need for Cognition Survey: 
\begin{enumerate}
  \item I would prefer complex to simple problems. (on a 5 point Likert scale)
  \item I like to have the responsibility of handling a situation that requires a lot of thinking. (on a 5 point Likert scale)
  \item Thinking is not my idea of fun. (on a 5 point Likert scale\edit{, reverse coded}) 
  \item I would rather do something that requires little thought than something that is sure to challenge my thinking abilities. (on a 5 point Likert scale\edit{, reverse coded})
  \item I really enjoy a task that involves coming up with new solutions to problems. (on a 5 point Likert scale)
  \item I would prefer a task that is intellectual, difficult, and important to one that is somewhat important but does not require much thought. (on a 5 point Likert scale)
\end{enumerate}

\subsection{Models for each hypothesis}
\noindent We use Bayesian models for all our analyses. \edit{We fit random intercepts of participants and mazes, as opposed to random slopes, as our fixed effects of interest (AI Condition, task difficulty, and explanation condition) do not vary within a group. We do account for random effects such as participant and maze.} We ran post-hoc tests \edit{using the emmeans package~\cite{emmeans}} on all models to find our specific effects. 

\noindent\textsc{H1a, H1b, h1c}:
% \begin{align*}
$\textrm{overreliance} \sim 1 + \textrm{Task difficulty} * \textrm{AI Condition} + (1 \mid \textrm{participant}) + (1 \mid \textrm{maze})$
% \end{align*}

\noindent\textsc{h1d}:
% \begin{align*}
$\textrm{overreliance} \sim 1 + \textrm{AI Condition} + (1 \mid \textrm{participant}) + (1 \mid \textrm{maze})$
% \end{align*}

\noindent\textsc{h1e}:
% \begin{align*}
$\textrm{overreliance} \sim 1 + \textrm{AI Condition} * \textrm{Need For Cognition} + (1 \mid \textrm{participant}) + (1 \mid \textrm{maze})$
% \end{align*}

\noindent\textsc{h2a}:
% \begin{align*}
$\textrm{overreliance} \sim 1 + \textrm{Explanation Condition} + (1 \mid \textrm{participant}) + (1 \mid \textrm{maze})$
% \end{align*}

\noindent\textsc{h2b}:
% \begin{align*}
$\textrm{overreliance} \sim 1 + \textrm{Explanation Condition} + (1 \mid \textrm{participant}) + (1 \mid \textrm{maze})$
% \end{align*}

\noindent\textsc{h3a}:
% \begin{align*}
$\textrm{overreliance} \sim 1 + \textrm{Bonus Condition} + (1 \mid \textrm{participant}) + (1 \mid \textrm{maze})$
% \end{align*}

\noindent\textsc{h3b}:
% \begin{align*}
$\textrm{overreliance} \sim 1 + \textrm{AI Condition} + (1 \mid \textrm{participant}) + (1 \mid \textrm{maze})$
% \end{align*}

\noindent\textsc{h4a}:
% \begin{align*}
$\textrm{utility} \sim 1 + \textrm{Task Difficulty Condition} + (1 \mid \textrm{participant})$
% \end{align*}

\noindent\textsc{h5a}:
% \begin{align*}
$\textrm{utility} \sim 1 + \textrm{Explanation Condition} + (1 \mid \textrm{participant})$
% \end{align*}

\edit{\noindent\textsc{Study 3 (non-pre-registered)}:
$\textrm{overreliance} \sim 1 + \textrm{AI Condition} + (1 \mid \textrm{participant}) + (1 \mid \textrm{maze})$}

\subsection{Previous pre-registrations}
We previously pre-registered two studies\footnote{\href{https://osf.io/h9256}{osf.io/h9256} and \href{https://osf.io/5q7r6}{osf.io/5q7r6}.} that made similar hypotheses using a popular Wikipedia Question Answering Task~\cite{rajpurkar2016squad}. 

\subsubsection{Study Design}
We had a similar study design as outlined in our main paper. The main difference is that, instead of the maze task, we use the Wikipedia Question-Answering Task, where our easy condition was one paragraph and our hard condition was five paragraphs. Our explanation condition was in the form of ``perfect" inline highlights, where the answer to the model's prediction was highlighted in the text, as common in prior work~\cite{bansal2021does,lai2020chicago,lai2019human,feng2019ai}.

\subsubsection{Hypotheses and results}
Our specific hypotheses can be found in the preregistration links.

Our hypotheses with the prior pre-registration did not appear to have a large effect, so we ended the study early at a number of participants far lower than pre-registered; therefore, we do not report these values.

\subsubsection{Understanding the results}
We posit that our studies using the Wikipedia Question Answering Task were (1) not suitable for detecting task difficulty differences and (2) that we previously had a conceptual error in our hypotheses. The former may be beause modulating the paragraph length to five paragraphs does not incur any significant extra cognitive cost, as it is easy to skim all five paragraphs to find the relevant key words. We find evidence for this in our qualitative analysis. At least 7/17 $(\sim41.1\%)$ participants in the hard prediction condition wrote in the free-form text box (which asked people something along the lines of, "how did you use the AI?") that they skimmed, scanned, or searched; for the second study, at least 17/57 $(\sim29.8\%)$ participants in the hard prediction condition wrote something about skimming, scanning, or searching. This also indicated that the hard condition was too easy to do.
The conceptual error regarding the hypothesis is due to initially thinking that explanations increase cost in the easy task, where we finally settled on the framing that explanations do not decrease the cost in the easy task. This prompted our switch from frequentist to Bayesian Statistics, where we are able to have ``no difference'' hypotheses.

\end{document}